\documentclass[preprints,article,accept,pdftex,moreauthors]{Definitions/mdpi} 
\firstpage{1} 
\pubvolume{1}
\issuenum{1}
\articlenumber{0}
\pubyear{2025}
\copyrightyear{2025}
\datereceived{ } 
\daterevised{ } 
\dateaccepted{ } 
\datepublished{ } 
\hreflink{https://doi.org/} 

\usepackage{amssymb}
\usepackage{amsmath}
\usepackage{multicol}
\usepackage{multirow}
\usepackage{booktabs}
\usepackage{longtable}
\usepackage{geometry}

\newcommand{\STAB}[1]{\begin{tabular}{@{}c@{}}#1\end{tabular}}
\Title{Recent advances in data-driven methods for degradation modelling across applications}

\TitleCitation{Recent advances in data-driven methods for degradation modelling across applications}

\Author{Anna Jarosz-Kozyro \orcidA{}, Jerzy Baranowski\orcidB{}}

\AuthorNames{Anna Jarosz-Kozyro, Jerzy Baranowski}

\isAPAStyle{%
       \AuthorCitation{Jarosz-Kozyro, A., \& Baranowski, J.}
         }{%
        \isChicagoStyle{%
        \AuthorCitation{Jarosz-Kozyro, Anna, and Jerzy Baranowski}
        }{
        \AuthorCitation{Jarosz-Kozyro, A.; Baranowski, J.}
        }
}

\address{Department of Automatic Control \& Robotics, AGH University of Kraków, 30-059 Kraków, Poland;\\
anjarosz@agh.edu.pl (A.J-K.);
jb@agh.edu.pl (J.B.)
}
\corres{Correspondence: {  jb@agh.edu.pl}}

\abstract{Understanding degradation is crucial for ensuring the longevity and performance of materials, systems, and organisms. To illustrate the similarities across applications, this article provides a review of data-based method in materials science, engineering, and medicine. The methods analyzed in this paper include regression analysis, factor analysis, cluster analysis, Markov Chain Monte Carlo, Bayesian statistics, hidden Markov models, nonparametric Bayesian modeling of time series, supervised learning, and deep learning. The review provides an overview of degradation models, referencing books and methods, and includes detailed tables highlighting the applications and insights offered in medicine, power engineering, and material science. It also discusses the classification of methods, emphasizing statistical inference, dynamic prediction, machine learning, and hybrid modeling techniques. Overall, this review enhances understanding of degradation modelling across diverse domains.}


\begin{document}

\section{Introduction}

Degradation is defined as a process in which the quality or condition of a material, system, or organism deteriorates over time, leading to a decline in performance or functionality. This process is crucial to understand for technical, engineering, and economical reasons, as it influences the service life of various structures and materials. In this article, we present a summary of existing research on degradation modeling, encompassing a diverse range of approaches across applications in material science, engineering, and medicine.

Degradation modeling plays a crucial role in various fields, including engineering, reliability analysis, and system optimization. Understanding the behavior of degrading systems and predicting their remaining useful life is essential for effective maintenance, resource allocation, and system design. In this review, we will present various statistical methods that have been applied to degradation modeling, exploring both traditional statistical techniques and more advanced approaches, such as machine learning algorithms. By examining these methods, we provide insights into the strengths, limitations, and potential applications of each approach in degradation modeling.

This paper is structured as follows. In Section 2 we introduce the key preliminaries on degradation modeling, including definitions, ISO standards across medicine, engineering and materials science, and the distinction between physical, data‐driven and knowledge‐based approaches. Section 3 surveys the existing literature, summarizing prior review papers and textbook treatments, and presents a comparative analysis of publication trends over time in the three application domains. In Section 4 we propose our classification of data-based methods into three main groups - statistical inference, dynamic prediction and machine learning - and describe each class in detail. Section 5 examines how these methods have been used in material science, power engineering and medicine, highlighting domain-specific preferences and hybrid combinations. Section 6 discusses open challenges and future research directions, such as uncertainty quantification, missing-data handling, data fusion, and scalability. Finally, Section 7 concludes with a summary of the main insights and recommendations for practitioners and researchers.

\section{Preliminaries}

\subsection{Degradation across applications}
A typical classification of degradation models discerns among three groups of models: physical models, data-driven models, and knowledge-based models \cite{Survey_Zagorowska2020}. Physical and knowledge-based models are typically application-specific because they describe particular phenomena occurring in a system. Conversely, data-based models focus on capturing the decline in performance or functionality regardless of the field of application. Despite the diverse applications of degradation, there are similarities in how degradation is described. Thus, similar data-based methods for degradation modelling can be used across disjoint applications. To illustrate the similarities across applications, this article provides a review of data-based method in materials science, engineering, and medicine.

In material science, degradation encompasses the deterioration of physical properties such as strength, ductility, and corrosion resistance of materials \cite{xia2020material}. Understanding degradation processes is crucial for selecting materials that can maintain their performance over time, ensuring the reliability and safety of structures and components. This includes various degradation mechanisms, such as fatigue, creep, oxidation, and environmental degradation, which impact the structural integrity and reliability of materials and components. 

In engineering, degradation refers to the decline in the performance or reliability of mechanical or electronic systems due to wear, fatigue, or aging \cite{Survey_Zagorowska2020}. This can include processes such as stress relaxation, thermal degradation, and chemical degradation, which affect the overall performance of materials and structures. Furthermore, the study of degradation in engineering involves the assessment of material degradation under different operational conditions, including temperature, humidity, and mechanical loading, to ensure the long-term functionality and safety of engineering systems.

In medicine, degradation focuses on the decline in the health or function of biological systems, such as tissues, organs, or physiological processes. This encompasses biodegradation and biodeterioration, where the vital activities of organisms lead to undesirable changes in the properties of materials. In addition, the understanding of degradation in medicine extends to biomaterial degradation, including implants corrosion, drug delivery systems degradation, and deterioration of biological tissues, with implications for the efficacy and safety of medical interventions.

Table \ref{tab:Definitions} summarizes the definitions of degradation as outlined in ISO standards across the fields of medicine, engineering, and material science. Each application has specific standards that address the implications and management of degradation in their respective contexts. 

\begin{table}[!tbp]
\begin{adjustwidth}{-0.1\extralength}{0cm}

\centering
\caption{Definitions of degradation from ISO Standards}
\label{tab:Definitions}
\begin{tabular}{p{2.1cm}p{12cm}}
\hline
\textbf{Application} & \textbf{Definitions from ISO Standards} \\ 
\hline
Medicine & ISO 10993-13:2010 -  the standard outlines methods for identifying and quantifying degradation products from polymeric medical devices, focusing on chemical alterations of the finished device \cite{ISO10993132010}. \\
& ISO 10993-1:2018 -  the standard provides a framework for the biological evaluation of medical devices, including considerations for degradation and its impact on biocompatibility \cite{ISO1099312018}. \\
& ISO 14971:2019 -  the standard addresses the application of risk management to medical devices, including risks associated with material degradation over time \cite{ISO149712019}. \\
& ISO 13485:2016 -  the standard specifies requirements for a quality management system where an organization needs to demonstrate its ability to provide medical devices that consistently meet customer and regulatory requirements, including those related to degradation \cite{ISO134852016}. \\  
Engineering                & ISO 55000:2014 -  the standard provides an overview of asset management, including the management of degradation in engineering components to ensure reliability and performance \cite{ISO550002014}. \\
& ISO 9001:2015 -  the standard outlines quality management principles that include monitoring and managing degradation in engineering processes and products \cite{ISO0012015}. \\
& ISO 14001:2015 -  the standard focuses on environmental management systems, which include considerations for material degradation and its environmental impacts \cite{ISO140012015}. \\
& ISO 50001:2018 -  the standard provides a framework for managing energy performance, which includes addressing degradation in materials and systems to improve energy efficiency \cite{ISO500012018}. \\  
Material Science           & ISO 15156-1:2015 -  the standard provides guidelines for materials used in oil and gas production, addressing degradation mechanisms such as corrosion and their impact on material selection \cite{ISO1515612015}. \\
& ISO 6892-1:2019 -  the standard specifies the method for tensile testing of metallic materials, which includes considerations for degradation effects on mechanical properties \cite{ISO689212019}. \\
& ISO 11469:2000 -  the standard provides guidelines for the identification of plastics and their degradation characteristics, focusing on environmental impacts \cite{ISO114692000}. \\
& ISO 14040:2006 -  the standard outlines principles and framework for life cycle assessment, which includes evaluating material degradation throughout the product life cycle \cite{ISO140402006}. \\ 
\hline
\end{tabular}
\label{tab:ISOreviews}
\end{adjustwidth}
\end{table}

Despite the specific manifestations of degradation in these fields, the common thread lies in understanding the degradation processes, identification of underlying degradation mechanisms, and prediction of future behavior. This shared objective underscores the interdisciplinary nature of degradation analysis and emphasizes the necessity of a cohesive approach to analyzing and modelling degradation.

\subsection{Data-based degradation modelling}

Mathematical models of degradation can be divided into analytical models, derived from first principles treating degradation as a physical phenomenon, and data-driven models, focused on describing degradation from experimental data \cite{habib2021data} without explicit knowledge about the underlying physics \cite{Solomatine2008}. Thus, developing models of degradation often relies on statistical inference to analyze degradation data, infer deterioration patterns, and provide reliability estimates and predictions based on historical data. Statistical inference methods consist in drawing conclusions about population parameters, making predictions, or testing hypotheses based on the estimated model. They often rely on model selection tools, such as the Akaike information criterion \cite{stoica2004model}, or employ model-free techniques that dynamically adapt to the contextual affinities of a process and capture intrinsic characteristics of the observations. 

Dynamic prediction methods model the dynamic evolution of degradation processes considering time-dependent changes in degradation mechanisms, and predicting future degradation behavior. This approach enables the assessment of degradation over time and the prediction of future performance based on dynamic changes in degradation mechanisms.

Lastly, the recent rise of machine learning techniques allows capturing intricate degradation patterns, identifying underlying degradation mechanisms, and making precise predictions using extensive and diverse datasets \cite{xu2021machine}. These approaches facilitate the comprehensive analysis of degradation behavior and the development of predictive models for various degradation processes.

\begin{figure}
    \centering
    \includegraphics[width=0.5\linewidth]{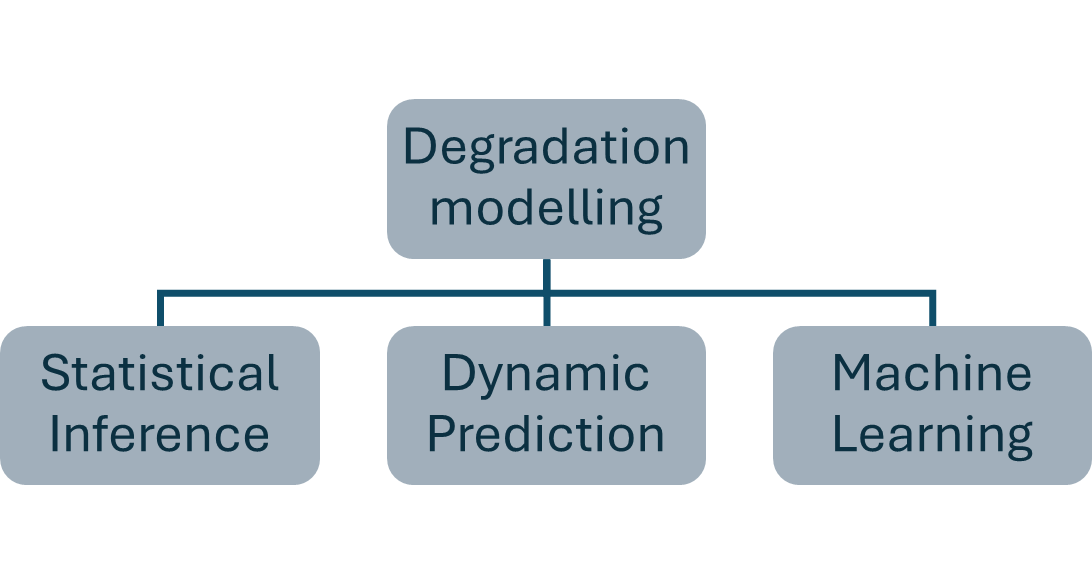}
    \caption{Main approaches to data-based modelling of degradation adopted in this work}
    \label{fig:ClassificationTreeSimple}
\end{figure}

The three approaches are shown in Fig. \ref{fig:ClassificationTreeSimple}.

\section{Background analysis}

\subsection{Existing reviews of data-based degradation modelling methods}
Table \ref{tbl:Books} provides an overview of methods for degradation modelling from textbooks, emphasizing the methods and proposed applications. The insights derived from these models contribute to the development of effective strategies for system maintenance, optimization, and reliability assessment. In particular, the books emphasize the fact that similar methods are used across applications. For instance, Bayesian modelling is discussed in \cite{Ghosal20171,Cardenas20181,Heard20211,Koch20071,Au20171,Barber20111} with application areas from economics and finance, through biostatistics and data analysis, to structural health monitoring. 

\begin{table}[!tbp]
\begin{adjustwidth}{-0.1\extralength}{0cm}

\caption{Degradation modelling in textbooks}
\label{tbl:Books}
\begin{tabular}{p{0.5cm}p{8cm}p{7cm}}
\hline
{Ref} & {Method} & {Application} \\ 
\hline
\cite{Ghosal20171} & regression analysis, Bayesian statistics  & CNC machines \\
\cite{Cardenas20181}  & Markov Chain Monte Carlo, Bayesian statistics & machining tools \\
\cite{Ceniga20211} & Markov Chain Monte Carlo, hidden Markov models & composite materials \\
\cite{Dong20201} & supervised learning, deep learning & Building materials \\
\cite{Jo20211} & cluster analysis, regression analysis &  industrial equipment \\
\cite{Gao20071} & nonparametric Bayesian modeling of time series, Bayesian statistics & infrastructure systems such as bridges or highways \\
\cite{Heard20211} & hidden Markov models, regression analysis & railway track geometry \\
\cite{Koch20071} & Markov Chain Monte Carlo, Bayesian statistics & industrial machinery \\
\cite{Hua20211} & Markov Chain Monte Carlo, supervised learning & brushless direct current motor \\
\cite{Baron20191} & hidden Markov models, nonparametric Bayesian modeling of time series & machinery under different stressors \\
\cite{Riguzzi20221} & Markov Chain Monte Carlo, Bayesian statistics &  metal components used in construction \\
\cite{Theodoridis20201} & hidden Markov models, nonparametric Bayesian modeling of time series & concrete structures \\
\cite{Au20171} & Bayesian statistics, regression analysis & structural components under stress \\
\cite{Barber20111} & nonparametric Bayesian modeling of time series, Markov Chain Monte Carlo & modal properties of structural systems \\
\cite{Brooks20111} & Markov Chain Monte Carlo, Bayesian statistics & engineering assets and materials \\
\hline
\end{tabular}
\end{adjustwidth}
\end{table}

Table \ref{tab:LitRev} summarizes existing literature reviews on degradation modeling. This compilation shows a broad spectrum of methodologies and classifications within the field of degradation modeling and provides an overview of the various approaches and classifications in the field of degradation modeling.

\begin{table}[!tbp]
\begin{adjustwidth}{-0.1\extralength}{0cm}

\caption{Literature reviews on degradation modeling}
\label{tab:LitRev}
\begin{tabular}{p{5cm}p{0.5cm}p{0.8cm}p{8cm}}
\hline
\textbf{Author} & \textbf{Ref} & \textbf{Year} & \textbf{Classification} \\ 
\hline
Firdaus, N., Ab-Samat, H., Prasetyo, B.T. & \cite{Firdaus2023640} & 2023 & defect detection model, Markovian model, machine learning-based predictive model \\
Jaime-Barquero, E., Bekaert, E., Olarte, J., Zulueta, E., Lopez-Guede, J.M. & \cite{JaimeBarquero2023} & 2023 & accelerated life testing model, physical-based model, machine Learning-based model
\\
Alimi, O.A., Meyer, E.L., Olayiwola, O.I. & \cite{Alimi2022} & 2022 & manual visual assessment model, condition monitoring model, statistical data analysis model \\
Berghout, T., Benbouzid, M. & \cite{Berghout2022} & 2022 & supervised learning model, unsupervised learning model, deep learning model \\
Zhao, S., Tayyebi, M., Mahdireza Yarigarravesh, Hu, G. & \cite{Zhao202312158} & 2023 &  mechanistic model, stochastic model, statistical model \\
Xue, K., Yang, J., Yang, M., Wang, D. & \cite{Xue2023} & 2023 & machine learning model, statistical model, data-driven model \\
Papargyri, L., Theristis, M., Kubicek, B., Papanastasiou, P., Georghiou, G.E. & \cite{Papargyri20202387} & 2020 & statistical model, machine learning model, simulation model \\
Mondal, M., Kumbhar, G.B. & \cite{Mondal2018483} & 2018 & neural network-based model, Monte Carlo simulation model, time series forecasting model \\
Zhang, M., Yang, S. & \cite{Zhang2024333} & 2024 & support vector clustering model, deep learning model,  statistical model \\
Chakurkar, P.S., Vora, D., Patil, S., Mishra, S., Kotecha, K. & \cite{Chakurkar2023} & 2023 & anomaly detection model, condition monitoring model,  time-series analysis model \\ 
\hline
\end{tabular}
\label{tab:degradation_models}
\end{adjustwidth}
\end{table}

\subsection{Comparative analysis across years}

\begin{figure}
    \centering
    \includegraphics[width=1\linewidth]{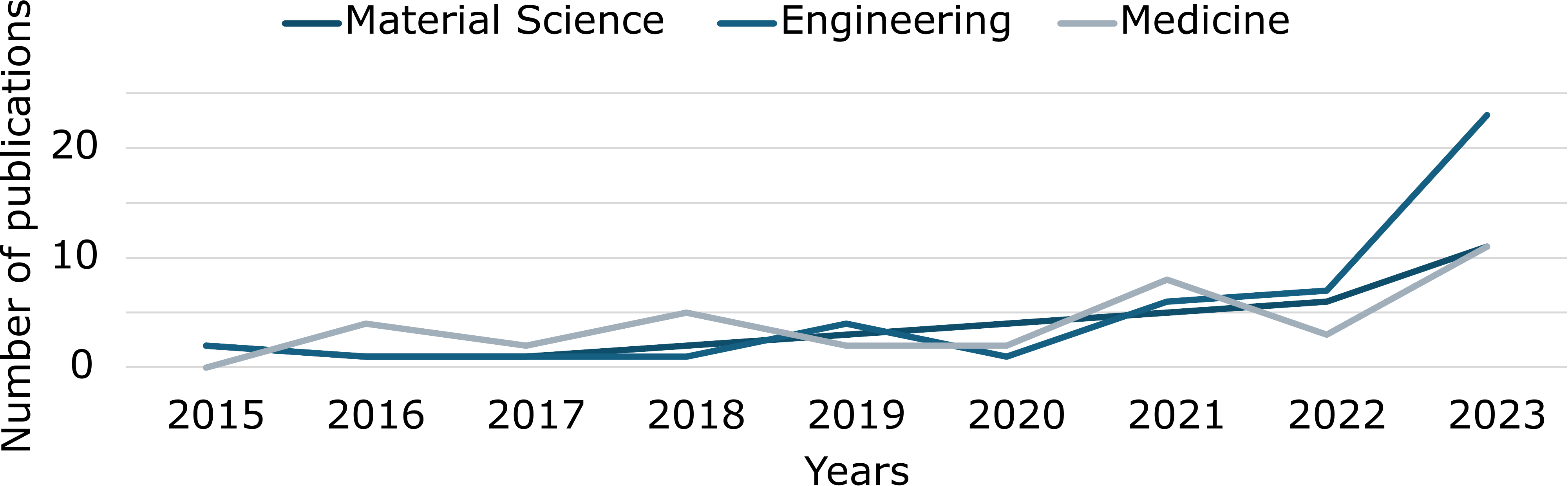}
    \caption{Number of articles over time in the three domains}
    \label{fig:Years}
\end{figure}

Figure \ref{fig:Years} shows the trends and variations in the numbers of articles across domains and years. In general, we observe an increasing trend in the number of articles over the years for all three domains. The largest increase is observed in Engineering, from two publication in 2015 to 23 publications in 2023. This increase may be explained by increased adoption of computational tools in engineering that allowed use of data-driven methods.

\section{Classification with respect to methods}

\begin{figure}
    \centering
    \includegraphics[width=0.55\linewidth]{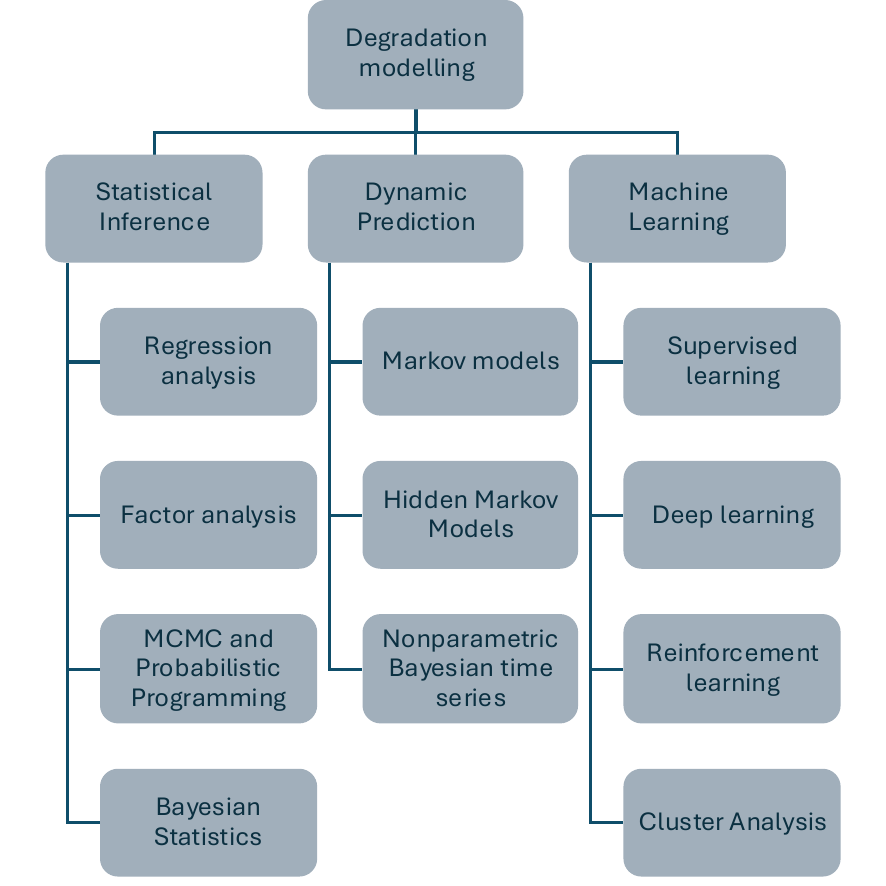}
    \caption{Classification tree of degradation modelling methods, adopted in this paper}
    \label{fig:ClassificationTree}
\end{figure}

This section presents methods used for degradation modelling. The classification of methods adopted in the paper is shown in Figure \ref{fig:ClassificationTree}.

\subsection{Statistical Inference}

Statistical inference models are used to simulate and analyze the processes of tissue and organ degradation in human organisms \cite{LI2018914}. These models allow for the analysis of the impact of factors such as aging, injuries, diseases, or the effect of drugs on the degradation of tissues and organs.

Statistical inference models are used to simulate and analyze technological processes in industry \cite{KurzR200170}. They are employed to forecast the behavior of technological systems, optimize production processes, and identify potential problems.

Statistical inference models are utilized to simulate and analyze various aspects related to the production, transmission, and use of energy \cite{Ceniga20211}. They are used to forecast energy consumption, optimize energy processes, and identify potential areas for energy efficiency improvement.

\subsubsection{Regression Analysis}
Regression analysis is a fundamental statistical tool for modeling the relationship between a dependent (response) variable and one or more independent (predictor) variables. In its simplest form, the linear regression model can be written as
\begin{equation}
    y_i = \beta_0 + \beta_1 x_{i1} + \cdots + \beta_p x_{ip} + \varepsilon_i,\quad i=1,\dots,n,
\end{equation}
where \(y_i\) denotes the observed response, \(x_{ij}\) the value of the \(j\)th predictor for the \(i\)th observation, \(\beta_0,\dots,\beta_p\) are unknown parameters, and \(\varepsilon_i\) are random errors, typically assumed to satisfy
\(\mathbb{E}[\varepsilon_i]=0\), \(\operatorname{Var}(\varepsilon_i)=\sigma^2\), and \(\varepsilon_i\perp\!\!\!\perp\varepsilon_j\) for \(i\neq j\).

Under these assumptions, the ordinary least squares (OLS) estimator for the parameter vector \(\boldsymbol\beta = (\beta_0,\beta_1,\dots,\beta_p)^\top\) is given in matrix form by
\begin{equation}
    \widehat{\boldsymbol\beta} = \bigl(\mathbf{X}^\top\mathbf{X}\bigr)^{-1}\mathbf{X}^\top\mathbf{y},
\end{equation}
where \(\mathbf{X}\in\mathbb{R}^{n\times(p+1)}\) is the design matrix whose first column is all ones and \(\mathbf{y}\in\mathbb{R}^n\) the vector of responses. The OLS solution minimizes the residual sum of squares,
\[
    \widehat{\boldsymbol\beta} = \arg\min_{\boldsymbol\beta}\,\|\mathbf{y} - \mathbf{X}\boldsymbol\beta\|^2.
\]
Regression analysis allows for hypothesis testing (e.g.\ \(H_0\colon \beta_j=0\)) and confidence interval estimation for coefficients, facilitating interpretability and inference. Extensions include weighted and generalized linear models to accommodate heteroscedasticity or non‐Gaussian responses \cite{montgomery2012introduction}.

\vspace{0.5\baselineskip}
\noindent\textbf{Key advantages:} simplicity, interpretability, and closed‐form solutions.  
\\
\noindent\textbf{Limitations:} reliance on linearity, independence, and homoscedasticity assumptions.

\noindent\textbf{Usage:} In the power industry, regression analysis precisely describes degradation processes by identifying the relationship between process variables and degradation \cite{Yu2021}. This allows for the rapid prediction of degradation based on known process parameters, optimizing maintenance activities.

Regression analysis is also used in medicine to mathematically describe degradation processes in organisms, such as the breakdown of chemicals in tissues \cite{Bejaoui2020703}. By employing regression analysis, researchers can determine the dynamics of degradation of medicinal substances in patients' bodies.

\subsubsection{Factor Analysis}

Factor analysis is a multivariate technique used to explain the covariance structure among a set of observed variables \(\mathbf{x} = (x_1,\dots,x_p)^\top\) in terms of a smaller number of unobserved latent factors \(\mathbf{f} = (f_1,\dots,f_m)^\top\), with \(m < p\). The classic common‐factor model is:
\begin{equation}
    \mathbf{x} = \boldsymbol{\Lambda}\,\mathbf{f} + \boldsymbol{\varepsilon},
\end{equation}
where \(\boldsymbol{\Lambda}\in\mathbb{R}^{p\times m}\) is the matrix of factor loadings, and \(\boldsymbol{\varepsilon}\sim N(\mathbf{0},\,\boldsymbol{\Psi})\) represents unique variances (specific plus error), with \(\boldsymbol{\Psi}\) diagonal. The covariance structure implied by the model is
\begin{equation}
    \operatorname{Cov}(\mathbf{x}) = \boldsymbol{\Sigma} = \boldsymbol{\Lambda}\boldsymbol{\Lambda}^\top + \boldsymbol{\Psi}.
\end{equation}
Estimating \(\boldsymbol{\Lambda}\) and \(\boldsymbol{\Psi}\) can be done via:
\begin{itemize}
  \item \emph{Principal axis factoring}, which iteratively estimates communalities \(h_j^2 = \sum_{k=1}^m \lambda_{jk}^2\) and solves eigenvalue problems on the reduced correlation matrix.
  \item \emph{Maximum likelihood (ML)}, which finds estimates \(\widehat{\boldsymbol{\Lambda}},\widehat{\boldsymbol{\Psi}}\) by maximizing the Gaussian log‐likelihood under \(H_0\colon \boldsymbol{\Sigma}(\boldsymbol{\Lambda},\boldsymbol{\Psi}) = \mathbf{S}\), where \(\mathbf{S}\) is the sample covariance.
\end{itemize}
Interpretation hinges on \emph{rotated} solutions (e.g.\ varimax) to achieve “simple structure” and on assessing model fit via likelihood‐ratio tests or information criteria.  

\vspace{0.5\baselineskip}

\noindent\textbf{Key advantages:} dimensionality reduction, uncovering latent constructs, and parsimonious modeling.  
\\
\noindent\textbf{Limitations:} identifiability issues, sensitivity to distributional assumptions, and subjective choice of number of factors \cite{bartholomew2011latent}.

\noindent\textbf{Usage:} In engineering processes, factor analysis can be used to identify groups of process variables \cite{Xu2016153}. They have a significant impact on degradation, leading to a better understanding of the degradation process.

Similarly, in the energy sector, factor analysis is employed to identify significant factors influencing degradation \cite{Lu2021354}. It helps optimize maintenance activities and improve understanding of degradation processes.

\subsubsection{Bayesian Statistics}
Bayesian statistics provides a principled framework for learning about unknown parameters \(\theta\) by combining prior beliefs \(p(\theta)\) with evidence from data \(\mathcal{D}\) through the likelihood \(p(\mathcal{D}\mid\theta)\).  The cornerstone is Bayes’ theorem:
\begin{equation}
    p(\theta\mid\mathcal{D}) \;=\; \frac{p(\mathcal{D}\mid\theta)\,p(\theta)}{p(\mathcal{D})}
    \quad\text{where}\quad
    p(\mathcal{D}) = \int p(\mathcal{D}\mid\theta)\,p(\theta)\,d\theta.
\end{equation}
The posterior \(p(\theta\mid\mathcal{D})\) quantifies updated uncertainty about \(\theta\).  From it one can derive:
\begin{itemize}
  \item \emph{Point estimates}, e.g.\ the posterior mean \(\mathbb{E}[\theta\mid\mathcal{D}]\).
  \item \emph{Credible intervals}, defined by
  \(\displaystyle \int_{a}^{b} p(\theta\mid\mathcal{D})\,d\theta = 1-\alpha.\)
  \item \emph{Posterior predictive distribution} for a new observation \(\tilde{y}\):
  \[
    p(\tilde{y}\mid\mathcal{D}) \;=\; \int p(\tilde{y}\mid\theta)\,p(\theta\mid\mathcal{D})\,d\theta.
  \]
\end{itemize}
Bayesian methods naturally handle hierarchical models by placing priors at multiple levels (e.g.\ \(\theta\sim p(\theta\mid\phi)\), \(\phi\sim p(\phi)\)), and they explicitly propagate all sources of uncertainty.  Computational tools such as Markov Chain Monte Carlo (MCMC) or variational inference allow approximation of posteriors when closed‐form solutions are unavailable \cite{gelman2013bayesian}.

\vspace{0.5\baselineskip}

\noindent\textbf{Key advantages:} coherent uncertainty quantification, flexible modeling of complex structures, and direct probability statements about parameters.  
\\
\noindent\textbf{Limitations:} computational intensity, sensitivity to prior choices, and challenges in high‐dimensional settings.

\noindent\textbf{Usage:} In the context of dynamic degradation processes of biologically active substances, Bayesian statistics is used to update knowledge based on clinical trial results \cite{Dadash2023}. This allows for the consideration of the variability of patients' physiological parameters and their impact on drug degradation.

Bayesian statistics updates knowledge about the degradation of biologically active substances based on clinical trial results \cite{Shahraki2017299}. It enhances the modeling of dynamic degradation processes.

\subsubsection{Markov Chain Monte Carlo and Probabilistic Programming}
Markov Chain Monte Carlo (MCMC) comprises a class of algorithms for generating samples from a target probability distribution \(\pi(\theta)\) when direct sampling is infeasible. A key property is that the Markov transition kernel \(K(\theta'\mid\theta)\) satisfies the \emph{detailed balance} condition
\[
\pi(\theta)\,K(\theta'\mid\theta)\;=\;\pi(\theta')\,K(\theta\mid\theta')
\quad\Longrightarrow\quad 
\int \pi(\theta)\,K(\theta'\mid\theta)\,d\theta = \pi(\theta').
\]
In the Metropolis–Hastings algorithm, one proposes \(\theta^*\sim q(\theta^*\mid\theta^{(t)})\) and accepts it with probability
\[
\alpha = \min\Bigl(1,\;\frac{\pi(\theta^*)\,q(\theta^{(t)}\mid\theta^*)}{\pi(\theta^{(t)})\,q(\theta^*\mid\theta^{(t)})}\Bigr),
\]
otherwise setting \(\theta^{(t+1)}=\theta^{(t)}\). A special case is the Gibbs sampler, which updates each component \(\theta_i\) by drawing from its full conditional \(\pi(\theta_i\mid\theta_{-i})\).

Hamiltonian Monte Carlo (HMC) augments the parameter space with auxiliary momentum variables \(p\in\mathbb{R}^d\) and uses simulated Hamiltonian dynamics to propose distant states with high acceptance probability \cite{neal2011mcmc}. One defines the Hamiltonian
\[
H(\theta,p) = U(\theta) + K(p) 
= -\log\pi(\theta) + \tfrac12\,p^\top M^{-1}p,
\]
where \(M\) is a mass matrix. Trajectories are simulated via the \emph{leapfrog} integrator:
\begin{align*}
p\Bigl(t+\tfrac{\epsilon}{2}\Bigr) &= p(t) - \tfrac{\epsilon}{2}\,\nabla_\theta U\bigl(\theta(t)\bigr),\\
\theta(t+\epsilon) &= \theta(t) + \epsilon\,M^{-1}p\Bigl(t+\tfrac{\epsilon}{2}\Bigr),\\
p(t+\epsilon) &= p\Bigl(t+\tfrac{\epsilon}{2}\Bigr) - \tfrac{\epsilon}{2}\,\nabla_\theta U\bigl(\theta(t+\epsilon)\bigr).
\end{align*}
After \(L\) steps, the proposal \((\theta^*,p^*)\) is accepted with probability \(\min\{1,\exp(-\Delta H)\}\).

Probabilistic programming languages (PPLs) such as Stan or BUGS allow users to specify complex hierarchical or latent‐variable models in high‐level syntax; the PPL runtime then automatically constructs and runs efficient MCMC chains (including HMC) to approximate posterior distributions of all parameters.

\vspace{0.5\baselineskip}

\noindent\textbf{Key advantages:} flexibility to model arbitrary posteriors, efficient exploration in high dimensions (HMC), and automatic uncertainty quantification.  
\\
\noindent\textbf{Limitations:} convergence diagnostics required, choice of integrator step‐size and path length in HMC, and potentially high computational cost.

\noindent\textbf{Usage:} In the context of variable energy conditions, MCMC accounts for uncertainties in modeling the degradation of biologically active substances \cite{Heard20211}. It enables effective prediction of future degradation states based on previous observations.

MCMC efficiently samples complex probability distributions, which is crucial in modeling step degradation \cite{Tamssaouet2023524}. It facilitates effective prediction of future degradation states based on previous observations.

\subsection{Dynamic Prediction}

In material science, Markov models enable engineers to analyze temporal dependencies and transitions between different states in a process. This allows for the prediction of future states, identification of bottlenecks, and optimization of process parameters for improved efficiency and productivity. Hidden Markov Models can uncover hidden patterns and detect anomalies in processes, aiding in process control and optimization.

In power engineering, dynamic prediction methods play a vital role in forecasting electricity demand, optimizing power generation, and managing energy resources. Markov models can accurately forecast short-term and long-term load demand by modeling the temporal dependencies and transitions in power consumption patterns. Hidden Markov Models help identify hidden states and patterns in power generation and consumption, facilitating efficient energy management and resource allocation. Nonparametric Bayesian time series modeling considers factors such as weather conditions and time of day to predict electricity demand accurately.

In medicine, dynamic prediction methods have significant applications, ranging from disease prognosis to treatment optimization. Markov models can model disease progression and predict patient outcomes based on current health states. Hidden Markov Models are valuable for modeling latent variables and hidden dynamics in medical data, enabling early detection of diseases and personalized treatment strategies. Nonparametric Bayesian time series modeling captures the complex temporal dynamics of patient data, facilitating accurate predictions of disease progression, treatment response, and prognosis.
\subsubsection{Markov models}

Markov models are useful for modeling abrupt degradation by considering the nonlinear relationships between process variables and degradation \cite{Au20171}. They enable the forecasting of future degradation states based on previous observations and can incorporate dynamic changes in the degradation process.

A (discrete‐time) Markov model represents the evolution of a system through a finite (or countable) set of states \(\mathcal{S}=\{1,\dots,K\}\), where the probability of a transition depends only on the current state.  If \(X_t\in\mathcal{S}\) denotes the state at time \(t\), the \emph{Markov property} is
\begin{equation}
  \Pr\bigl(X_{t+1}=j \mid X_t=i,\,X_{t-1},\dots,X_0\bigr)
  \;=\;
  \Pr\bigl(X_{t+1}=j \mid X_t=i\bigr)
  =: p_{ij},\quad i,j\in\mathcal{S}.
\end{equation}
The one‐step transition probabilities \(p_{ij}\) form the transition matrix \(P=(p_{ij})\), satisfying \(\sum_{j}p_{ij}=1\) for each \(i\).  The \(n\)‐step transition probabilities are given by the Chapman–Kolmogorov equations,
\begin{equation}
  p_{ij}^{(n)} \;=\; \Pr(X_{t+n}=j \mid X_t=i)
  \;=\;
  \bigl(P^n\bigr)_{ij},\quad
  P^n = \underbrace{P\cdots P}_{n\text{ times}}.
\end{equation}
If \(\boldsymbol\pi^{(t)}\) is the row vector of state‐probabilities at time \(t\), then
\[
  \boldsymbol\pi^{(t+n)} \;=\; \boldsymbol\pi^{(t)}\,P^n.
\]
For ergodic chains (irreducible and aperiodic), there exists a unique stationary distribution \(\boldsymbol\pi^*\) satisfying \(\boldsymbol\pi^* = \boldsymbol\pi^*\,P\); this can be used for long‐term degradation prediction.  In absorbing chains, mean time to absorption and absorption probabilities provide estimates of remaining useful life.

\vspace{0.5\baselineskip}

\noindent\textbf{Key advantages:} simple formulation, closed‐form \(n\)‐step predictions, and well‐studied theory of long‐run behavior.  
\\
\noindent\textbf{Limitations:} state‐space discretization may be coarse, and the loss of history beyond the current state may oversimplify gradual degradation \cite{norris1997markov}.

\noindent\textbf{Usage:} In the context of discontinuous energy processes, Markov models are important for modeling abrupt degradation and considering dynamic changes in the degradation process, especially under changing energy conditions \cite{Zhou2009270}. They also allow for the inclusion of nonlinear relationships between process variables and degradation.

In medicine, Markov models can be used to model changes in patients' health status, predict the course of diseases, and assess the risk of drug degradation \cite{Bulinski2017639}. They consider the dynamic changes in patients' bodies and their impact on drug degradation.

\subsubsection{Hidden Markov Models}

Hidden Markov Models are used to model abrupt degradation by considering the nonlinear relationships between process variables and degradation \cite{Zhang2010743}. They enable the forecasting of future degradation states based on previous observations.

A Hidden Markov Model (HMM) describes a system where an unobserved (hidden) state sequence \(\{X_t\}\) evolves as a Markov chain over states \(\mathcal{S}=\{1,\dots,K\}\), and at each time \(t\) emits an observation \(Y_t\) according to a state‐dependent distribution.  An HMM is specified by:
\begin{itemize}
  \item Initial distribution \(\pi_i = \Pr(X_1=i)\).
  \item Transition matrix \(A = (a_{ij})\) with \(a_{ij} = \Pr(X_{t+1}=j\mid X_t=i)\).
  \item Emission probabilities \(b_i(y) = \Pr(Y_t=y\mid X_t=i)\) (or density \(b_i(y)\) for continuous \(Y_t\)).
\end{itemize}
The joint probability of a state sequence \(\mathbf{x}=(x_1,\dots,x_T)\) and observations \(\mathbf{y}=(y_1,\dots,y_T)\) is
\[
  \Pr(\mathbf{x},\mathbf{y})
  = \pi_{x_1}\,b_{x_1}(y_1)\,\prod_{t=2}^T a_{x_{t-1},x_t}\,b_{x_t}(y_t).
\]
Key computations include:  
\begin{enumerate}
  \item \emph{Filtering / likelihood}: via the forward recursion  
    \(\alpha_t(j)=\bigl[\sum_i\alpha_{t-1}(i)a_{ij}\bigr]\,b_j(y_t)\).  
  \item \emph{Decoding}: the Viterbi algorithm finds \(\arg\max_{\mathbf{x}}\Pr(\mathbf{x}\mid\mathbf{y})\).  
  \item \emph{Learning}: Baum–Welch (EM) updates \(\pi,A,b\) by maximizing the data likelihood.
\end{enumerate}
HMMs capture abrupt or regime‐switching degradation by modeling latent condition changes and corresponding observation patterns.  

\vspace{0.5\baselineskip}

\noindent\textbf{Key advantages:} ability to model unobserved regimes, efficient inference via dynamic programming.  
\\
\noindent\textbf{Limitations:} choice of state‐space size, assumption of conditional independence of observations \cite{rabiner1989tutorial}.

\noindent\textbf{Usage:} In the context of discontinuous energy processes, Hidden Markov Models are important for modeling abrupt degradation \cite{Du2018}. It helps consider dynamic changes in the degradation process, especially under changing energy conditions.

Hidden Markov Models can be effectively used in medicine to model changes in patients' health, taking into account dynamic fluctuations in patients' bodies and their impact on drug degradation \cite{Baldi2017329}. These advanced models allow for predicting the effectiveness of therapies.

\subsubsection{Nonparametric Bayesian Time Series Modeling}
Nonparametric Bayesian methods allow the model complexity to grow with data by placing priors on infinite‐dimensional objects. Three widely used approaches in time series are:

\paragraph{Dirichlet Process Mixtures}  
Partition the series \(\{y_t\}_{t=1}^T\) into an unknown number of regimes, indexed by latent parameters \(\{\theta_k\}\). A Dirichlet Process (DP) prior on the mixing measure \(G\) induces a countably infinite mixture:
\begin{align}
  G &\sim \mathrm{DP}(\alpha, G_0),\nonumber\\
  \theta_t \mid G &\sim G,\quad
  y_t \mid \theta_t \;\sim\; F\bigl(y_t \mid \theta_t\bigr).
\end{align}
Using the stick‐breaking construction,
\[
  G = \sum_{k=1}^\infty \pi_k\,\delta_{\theta_k},\quad
  \pi_k = \beta_k\prod_{l<k}(1-\beta_l),\quad
  \beta_k\sim\mathrm{Beta}(1,\alpha),
\]
one obtains an adaptive mixture that clusters observations into regimes with similar dynamics. Posterior inference (e.g.\ via Gibbs sampling) estimates both the number of regimes and their parameters.

\paragraph{Gaussian Process Regression}  
Model the latent signal \(f\colon t\mapsto\mathbb{R}\) as a Gaussian Process (GP):
\begin{align}
  f(t) &\sim \mathcal{GP}\bigl(m(t),\,k(t,t')\bigr),\nonumber\\
  y_t &= f(t) + \varepsilon_t,\quad
  \varepsilon_t \sim N(0,\sigma^2).
\end{align}
The kernel \(k(t,t')\) encodes smoothness or periodicity, and the GP posterior provides closed‐form predictive distributions:
\[
  f_* \mid \mathbf{y} \;\sim\;
  N\bigl(K_*K^{-1}\mathbf{y},\;K_{**} - K_*K^{-1}K_*^\top\bigr),
\]
where \(K = [k(t_i,t_j)]\), \(K_* = [k(t_*,t_j)]\), etc.

\paragraph{Prophet}  
Prophet is an additive forecasting model designed for business time series, decomposing \(y_t\) into trend \(g(t)\), seasonality \(s(t)\), holiday effects \(h(t)\), and noise:
\begin{equation}
  y_t = g(t) + s(t) + h(t) + \varepsilon_t,\quad \varepsilon_t\sim N(0,\sigma^2).
\end{equation}
The trend \(g(t)\) is modeled as a piecewise linear (or logistic) function with automatic changepoint detection:
\[
  g(t) = \bigl(k + a(t)^\top\boldsymbol\delta\bigr)\,t + \bigl(m + a(t)^\top\boldsymbol\gamma\bigr),
\]
where \(a(t)\) indicates segments, \(k,m\) are base rate and offset, and \(\boldsymbol\delta,\boldsymbol\gamma\) are adjustments at changepoints. Seasonality \(s(t)\) uses Fourier series:
\[
  s(t) = \sum_{n=1}^{N}\Bigl[\alpha_n\cos\bigl(2\pi n t/P\bigr) + \beta_n\sin\bigl(2\pi n t/P\bigr)\Bigr],
\]
and \(h(t)\) includes user‐specified events. Prophet is fitted via MAP estimation with weakly informative priors, yielding fast, interpretable forecasts \cite{taylor2018forecasting}.

\vspace{0.5\baselineskip}

\noindent\textbf{Key advantages:} automatic changepoint detection, interpretable components, and scalability to large datasets.  
\\
\noindent\textbf{Limitations:} assumes additive structure, may struggle with highly irregular dynamics, and limited probabilistic uncertainty beyond the MAP fit.

\subsection{Machine Learning}

Machine learning methods enable automatic detection of patterns in process data and building predictive models based on them \cite{Ding20191272}. They facilitate adaptive degradation modeling, allowing for the modeling of degradation in engineering processes using a wide range of predictive algorithms.

In the power industry, machine learning methods enable adaptive modeling of degradation \cite{Zhang20121}. It may consider changing process conditions and optimizing maintenance operations.

In medicine, machine learning methods can be used to predictively model drug degradation based on clinical and laboratory data \cite{Jo20211}. They enable adaptive modeling of drug degradation, automatic detection of drug degradation patterns, and optimization of therapeutic doses.

\subsubsection{Supervised Learning}

Supervised learning aims to learn a mapping \(f:\mathcal{X}\to\mathcal{Y}\) from input–output pairs \(\mathcal{D}=\{(x_i,y_i)\}_{i=1}^n\), where each \(y_i\) is a known label or response.  The model \(f(x;\theta)\), parameterized by \(\theta\), is trained by minimizing an empirical risk functional:
\begin{equation}
  \widehat{\theta} \;=\;\arg\min_{\theta}\;\frac{1}{n}\sum_{i=1}^n \ell\bigl(y_i,\,f(x_i;\theta)\bigr)
\end{equation}
where \(\ell(\cdot,\cdot)\) is a loss function (e.g.\ squared error \(\ell(y,\hat y)=(y-\hat y)^2\) for regression or cross‐entropy \(\ell(y,\hat y)=-\sum_k y_k\log \hat y_k\) for classification).  

Two primary supervised tasks are:
\begin{itemize}
  \item \emph{Regression}: \(\mathcal{Y}=\mathbb{R}\), predicting continuous degradation measures (e.g.\ wear rate).
  \item \emph{Classification}: \(\mathcal{Y}=\{1,\dots,C\}\), labeling discrete states (e.g.\ “healthy” vs.\ “faulty”).
\end{itemize}
Common algorithmic frameworks include linear models, support vector machines, decision trees and ensembles, and feed‐forward neural networks.  Model selection and regularization (e.g.\ ridge, lasso) control complexity and mitigate overfitting.

\vspace{0.5\baselineskip}

\noindent\textbf{Key advantages:} direct use of labeled data, flexibility across tasks, and mature theory for generalization (e.g.\ VC‐dimension, Rademacher complexity).  
\\
\noindent\textbf{Limitations:} requires substantial labeled data, sensitive to label noise, and potential overfitting without proper regularization \cite{hastie2009elements}.

\noindent\textbf{Usage:} In the power industry, supervised learning methods enable adaptive modeling of degradation \cite{Qin2017}. It considers changing process conditions, and optimizing maintenance operations.

In medicine, supervised learning allows for adaptive modeling of drug degradation based on individual patient characteristics \cite{Kang20201137}. It may optimize therapeutic doses and assessing degradation risk.

\subsubsection{Deep Learning}
Deep learning refers to a class of machine learning methods based on artificial neural networks with multiple hidden layers, which can automatically learn hierarchical feature representations from raw data.  A feed‐forward deep neural network with \(L\) layers can be defined recursively as
\begin{align}
  \mathbf{a}^{(0)} &= \mathbf{x},\nonumber\\
  \mathbf{z}^{(l)} &= W^{(l)}\mathbf{a}^{(l-1)} + \mathbf{b}^{(l)},\quad l=1,\dots,L,\nonumber\\
  \mathbf{a}^{(l)} &= \phi\bigl(\mathbf{z}^{(l)}\bigr),
\end{align}
where \(\mathbf{x}\in\mathbb{R}^d\) is the input, \(W^{(l)}\) and \(\mathbf{b}^{(l)}\) are the weights and biases of layer \(l\), and \(\phi(\cdot)\) is a nonlinear activation (e.g.\ ReLU, sigmoid).  The network’s output \(\hat{\mathbf{y}} = \mathbf{a}^{(L)}\) is compared to the true label \(\mathbf{y}\) via a loss function \(\ell(\mathbf{y},\hat{\mathbf{y}})\), and parameters are optimized by backpropagation and stochastic gradient descent:
\[
  \theta \leftarrow \theta - \eta \,\nabla_\theta \frac{1}{n}\sum_{i=1}^n \ell\bigl(\mathbf{y}_i,\,\hat{\mathbf{y}}_i\bigr).
\]
Extensions include convolutional neural networks (CNNs) for spatial degradation patterns and recurrent architectures (RNNs / LSTMs) for temporal degradation sequences.  Deep models excel at capturing complex, nonlinear relationships in high‐dimensional sensor or image data, which is critical for accurate degradation prediction and anomaly detection.

\vspace{0.5\baselineskip}

\noindent\textbf{Key advantages:} automatic feature learning, state‐of‐the‐art predictive performance in large datasets, and flexible architectures for diverse data modalities.  
\\
\noindent\textbf{Limitations:} large data and computational requirements, potential overfitting, and reduced interpretability of learned features \cite{goodfellow2016deep}.

\noindent\textbf{Usage:} 
In the field of power engineering, deep learning facilitates advanced pattern recognition in modeling degradation and predicting energy processes \cite{Dong20201}. It enables adaptive modeling of degradation risk in energy processes.

In medicine, deep learning enables advanced recognition of drug degradation patterns \cite{Lu2021354}. It contributes to adaptive modeling, prediction of therapy efficacy, and assessment of degradation risk.

\subsubsection{Reinforcement Learning}
Reinforcement learning (RL) addresses sequential decision–making under uncertainty by learning a policy \(\pi(a\mid s)\) that maximizes expected cumulative reward in a Markov decision process (MDP). An MDP is defined by the tuple \((\mathcal{S},\mathcal{A},P,r,\gamma)\) where:
\begin{itemize}
  \item \(\mathcal{S}\) is the state space, \(\mathcal{A}\) the action space.
  \item \(P(s' \mid s,a)\) is the transition probability.
  \item \(r(s,a)\) is the immediate reward.
  \item \(\gamma\in[0,1)\) is the discount factor.
\end{itemize}
The goal is to maximize the \emph{return} \(G_t = \sum_{k=0}^\infty \gamma^k\,r(s_{t+k},a_{t+k})\).  The state‐value and action‐value functions under policy \(\pi\) satisfy the Bellman equations:
\begin{align}
  V^\pi(s) &= \mathbb{E}_\pi\bigl[G_t \mid s_t=s\bigr]
           = \sum_{a}\pi(a\mid s)\Bigl[r(s,a) + \gamma\sum_{s'}P(s'\mid s,a)V^\pi(s')\Bigr],\\
  Q^\pi(s,a) &= r(s,a) + \gamma\sum_{s'}P(s'\mid s,a)\sum_{a'}\pi(a'\mid s')\,Q^\pi(s',a').
\end{align}
Model‐free methods include:
\begin{itemize}
  \item \emph{Q‐learning} (off‐policy):  
    \(\displaystyle Q_{t+1}(s_t,a_t) \leftarrow Q_t(s_t,a_t) + \alpha\bigl[r_t + \gamma\max_{a'}Q_t(s_{t+1},a') - Q_t(s_t,a_t)\bigr].\)
  \item \emph{Policy gradient} (on‐policy): optimize \(J(\theta)=\mathbb{E}_{\pi_\theta}[G_t]\) via  
    \(\nabla_\theta J = \mathbb{E}_{\pi_\theta}\bigl[\nabla_\theta\log\pi_\theta(a\mid s)\,Q^{\pi_\theta}(s,a)\bigr].\)
\end{itemize}
Deep RL integrates neural networks to approximate \(Q\) or \(\pi\) (e.g.\ DQN, actor–critic).  In degradation modeling, RL can learn maintenance or control policies that dynamically trade off immediate performance versus long‐term system health.

\vspace{0.5\baselineskip}

\noindent\textbf{Key advantages:} learns adaptive policies without explicit system models; handles stochastic, nonstationary environments.  
\\
\noindent\textbf{Limitations:} sample‐inefficient; high variance in gradient estimates; requires careful tuning of hyperparameters \cite{sutton2018reinforcement}.

\noindent\textbf{Usage:} 
In the field of power engineering, reinforcement learning is used for adaptive degradation modeling \cite{Samaranayake201889}. It considers dynamic changes in the degradation process, especially under changing energy conditions.

In medicine, reinforcement learning is employed for adaptive modeling of drug degradation \cite{Fang2020}. It enables the prediction of therapy effectiveness and the assessment of degradation risk.

\subsubsection{Cluster Analysis}
Cluster analysis is an unsupervised learning technique that seeks to partition a set of \(n\) observations \(\{x_i\in\mathbb{R}^d\}_{i=1}^n\) into \(K\) groups (clusters) \(\{C_k\}_{k=1}^K\) such that observations within the same cluster are more similar to each other than to those in other clusters.  Two common approaches are:

\paragraph{K‐Means Clustering}  
Assign each \(x_i\) to the cluster with the nearest centroid \(\mu_k\), by minimizing the within‐cluster sum of squares (WCSS):
\begin{equation}
  \{\widehat{C}_k,\widehat{\mu}_k\}_{k=1}^K
  = \arg\min_{\{C_k,\mu_k\}}\;\sum_{k=1}^K\sum_{x_i\in C_k}\|x_i-\mu_k\|^2.
\end{equation}
The standard algorithm alternates between (i) \emph{assignment}: \(C_k \leftarrow \{\,i:\|x_i-\mu_k\|\le\|x_i-\mu_j\|\;\forall j\}\), and (ii) \emph{update}: \(\mu_k \leftarrow \tfrac{1}{|C_k|}\sum_{x_i\in C_k}x_i\), until convergence.

\paragraph{Hierarchical Clustering}  
Builds a tree (dendrogram) of clusters either by agglomeration (bottom‐up) or division (top‐down).  In agglomerative clustering, start with each point as its own cluster and at each step merge the pair \((C_a,C_b)\) minimizing a linkage criterion, e.g.\  
\[
  d_{\text{single}}(C_a,C_b) = \min_{x_i\in C_a,x_j\in C_b}\|x_i-x_j\|,
  \quad
  d_{\text{complete}}(C_a,C_b) = \max_{x_i\in C_a,x_j\in C_b}\|x_i-x_j\|.
\]

\noindent By cutting the dendrogram at a chosen height, one obtains a desired number of clusters.

\vspace{0.5\baselineskip}

\noindent\textbf{Key advantages:} no need for labeled data; can discover unknown structure; interpretable via centroids or dendrograms.  
\\
\noindent\textbf{Limitations:} choice of \(K\) or cut‐height is subjective; 
sensitive to scaling and noise; may find only spherical clusters (for K‐means) or be computationally expensive (\(\mathcal{O}(n^3)\) for naive hierarchical implementations) \cite{kaufman2005finding}.

\noindent\textbf{Usage:} 
In the power industry, cluster analysis helps reduce data complexity by identifying important factors influencing degradation processes \cite{deLima2021}. It enables mathematical modeling of various degradation cases, contributing to a better understanding of degradation processes.

Cluster analysis is also used in medicine to group degradation events based on similarities in their characteristics \cite{Bejaoui2020703}. It helps identify groups of process variables that significantly impact the degradation of chemicals in organisms.

\section{Classification of methods and their usage across applications}
This section presents degradation modeling across domains such as Material Science, Engineering, and Medicine. In Table \ref{tab:methodsoverall}, we provide a comprehensive overview of the number of articles published in each domain, categorized by the methods employed. This analysis reveals the distribution of research efforts across the three domains, highlighting how each method contributes to the understanding and application of degradation modeling. 

\begin{table}[!tbp]
\centering
\caption{Method used for degradation modelling with dynamic prediction}
\label{tab:methodsoverall}
\begin{tabular}{p{3cm}p{3cm}p{3cm}p{3cm}}
\hline
Method & Material Science & Engineering & Medicine \\
\hline
Statistical inference & 25 & 41 & 34 \\
Dynamic prediction & 21 & 36 & 28 \\
Machine learning & 23 & 31 & 27 \\
\hline
\end{tabular}
\end{table}

\subsection{Individual methods}
\begin{figure}[!tbp]
{
    \centering
\subfloat[Machine learning]{
\includegraphics[width=0.8\linewidth]{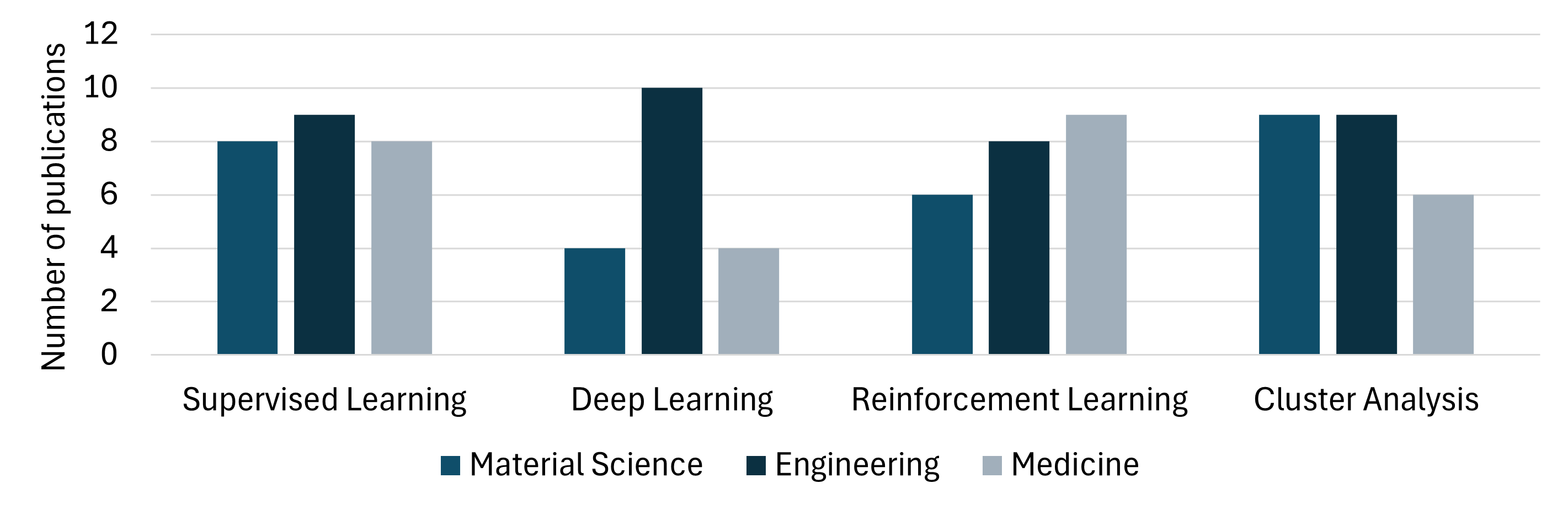}
    \label{fig:ML3}
}\\
\subfloat[Deep Learning]{
\includegraphics[width=0.8\linewidth]{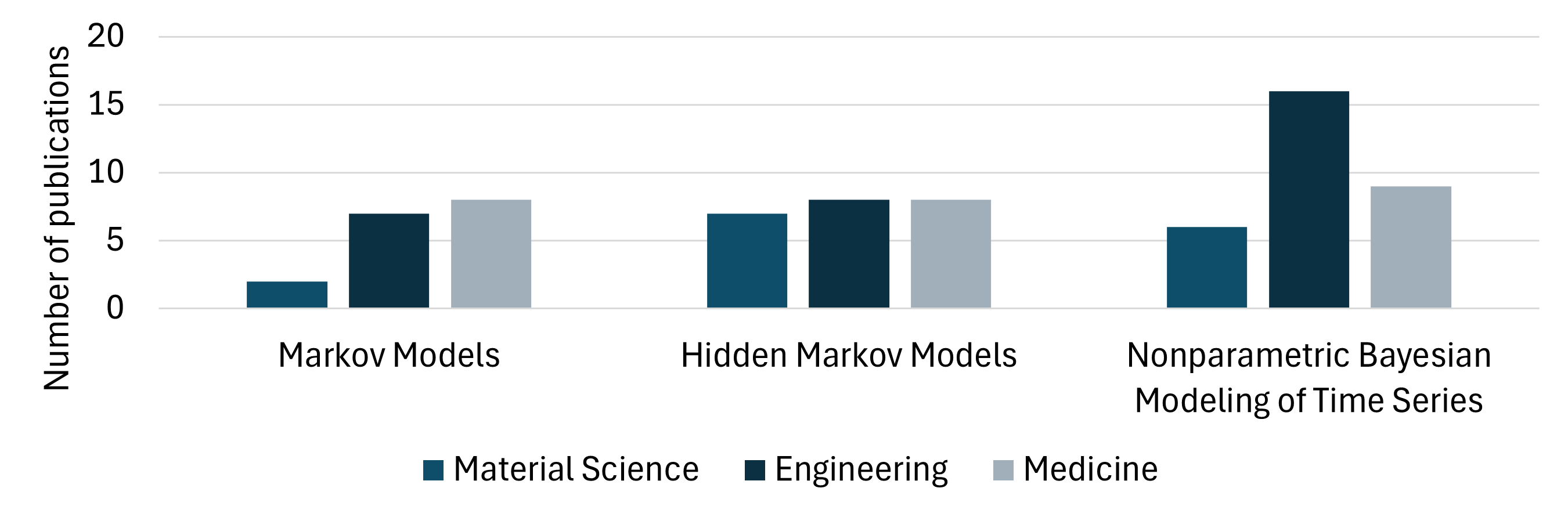}
    \label{fig:DP3}
}\\
\subfloat[Statistical Inference]{
\includegraphics[width=0.8\linewidth]{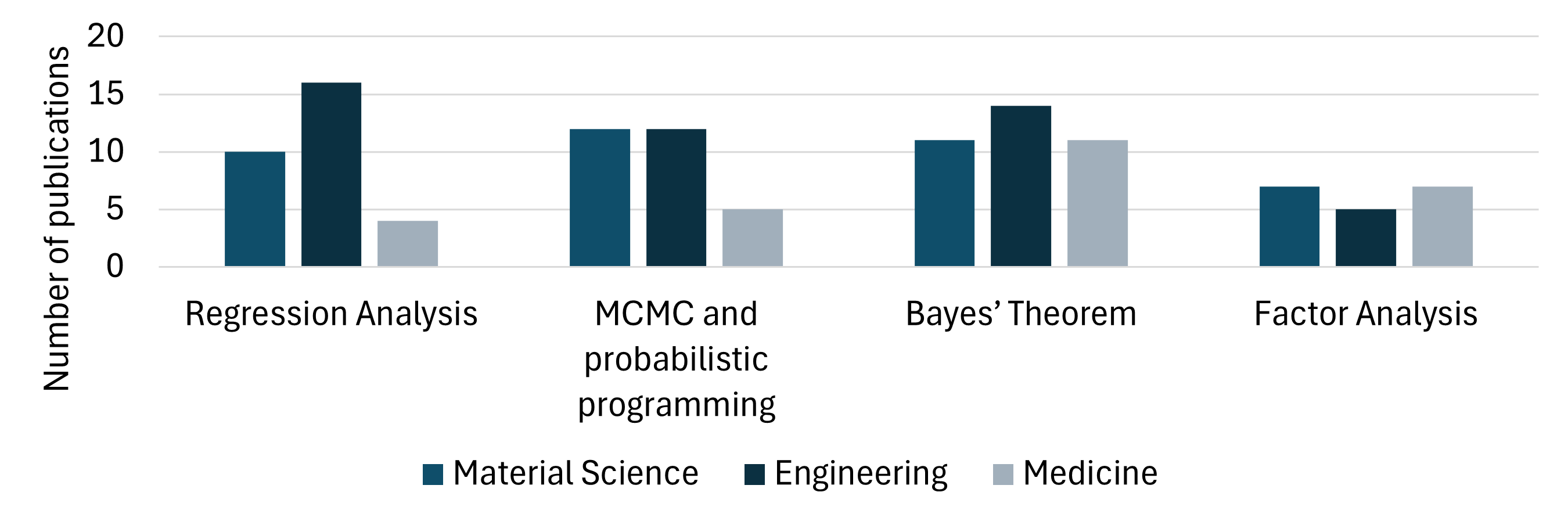}
    \label{fig:SI3}
}
    \caption{Usage of techniques}
\label{fig:Usage}
}
\end{figure}

Figure \ref{fig:SI3} provides an overview of the utilization of different techniques in the fields of material science, engineering, and medicine. It encompasses three main categories: statistical inference, dynamic prediction methods, and machine learning.

\subsubsection{Statistical inference usage}

Table \ref{tab:methodsSI} shows the use of techniques from the statistical inference method in three disciplines: material science, power engineering, and medicine.

Markov Chain Monte Carlo (MCMC) + probabilistic programming is applicable in all three disciplines: Material Science, Engineering, and Medicine. This technique is used for sampling from probability distributions and approximating the desired distribution.
Regression Analysis is more applicable in Engineering compared to Material Science and Medicine. It is a statistical technique used to model the relationship between variables and make predictions.
Bayesian statistics is applicable in all three disciplines, with higher relevance in Medicine. It is a fundamental concept in Bayesian inference and has applications in probabilistic modeling.
Factor Analysis has similar relevance in Material Science and Medicine, while being less applicable in Engineering. It is a statistical technique used to explore relationships among observed variables and identify underlying factors. 

Table \ref{tab:methodsSI} presents a compilation of articles related to degradation models with statistical inference in the field of Material Science. These articles explore various techniques used to study the degradation of materials and provide insights into their behavior under different conditions.

\begin{table}[htbp]
\centering
\caption{Method used for degradation modelling with statistical inference}
\label{tab:methodsSI}
\begin{tabular}{p{3cm}p{3cm}p{3cm}p{3cm}}
\hline
Method & Material Science & Engineering & Medicine \\
\hline
Regression Analysis & \cite{Jayakumar2021,Wu2024,RodriguezBravo2019394,Liu2023,Hosseini2018,ZellerPlumhoff20214368,AlMahdi2023604,Yang2022a,Alessio2020,Elahi202330}  & \cite{Li2024,Jain2024,Han2022,Osara2019,Fang202332884,Shchurov2021,Stroe2018517,Oria20236539,Karunathilake2023,Xu2015A2026,Liu2012336,Spitthoff2023,Su2023,Narale2019,Kohtz2024,Chaleshtori2024} & \cite{Dumanlidag2021246,Vernizzi2020,Wang202491,Wang2023} \\
Factor Analysis & \cite{Nguyen2024,Lin202137,AmayaGomez201926,Weng2023,Liu2022,Nepal20131060,Schneider2023} & \cite{Kulkarni2009,Chen2023a,Qu2022441,Reniers2019A3189,Nicolai2016153} & \cite{Godinho2021,Liang2022,Ewald2023,Sheng201867,Geng2023,Rashid2018,Song2016114} \\
Markov Chain Monte Carlo + probabilistic programming & \cite{Yang2022b,DUrso2024,Alessio2020,Zheng2024,Hu2023,Szabelski2022,Kaewmala2023,Meng2021,Schneider2023,Ueda2020,Silva2019,Chunsheng2017} & \cite{Park2024,Mitra2023,Li202239,Singh20172481,Schiffmacher2021895,Costa2022,Yu20061840,Zhao2023,Chen2023a,Streb2023,Yamaguchi2023,Spitthoff2023} & \cite{Snuggs2019,RiveroRios2016238,Ayala2018,Betancourt2019,Shuai2021490} \\
Bayesian statistics & \cite{Badet2020,Gupta2023,Chaudhary2023,Dalmau2018167,ZellerPlumhoff20214368,Xiang2011768,Kudo201531,AfsarDizaj20225374,Weng2023,Kropka201514,Li2022} & \cite{Shi2023,Lu2023,Su2024,Xu2015A2026,Zhao2024,Bai2023,Peng20212105,Erdinc2009383,Haque2023,Kathribail2023305,Yan2021,Jeong2023,Fan2021,Alam20135592,Liu202381} & \cite{Liang2022,Ewald2023,Dumanlidag2021246,Geng2023,Ohnishi2016,Tanino2023,Gredic2023,Lai2021,Zheng2023,Kong2008327,Hwang2017} \\
\hline
\end{tabular}
\end{table}

\subsubsection{Dynamic prediction usage}

Table \ref{tab:methodsDP} refers to the number of applications of individual techniques from the dynamic prediction methods method in each of the disciplines, which allows us to conclude the preferences and popularity of particular techniques in particular fields.

Markov Models and Hidden Markov Models are applicable in all three disciplines: Material Science, Engineering, and Medicine. These models are used to analyze and model sequential data with hidden states.
Nonparametric Bayesian Modeling of Time Series is more applicable in Engineering compared to Material Science and Medicine. This approach allows for flexible modeling of time series data without making strong assumptions about the underlying distribution.
The relevance of these techniques varies across disciplines. For example, Medicine shows higher relevance for Markov Models and Hidden Markov Models compared to Material Science and Engineering. This suggests that these techniques have specific applications or importance in the medical field.

\begin{table}[!tbp]
\centering
\caption{Method used for degradation modelling with Dynamic Prediction}
\label{tab:methodsDP}
\begin{tabular}{p{3cm}p{3cm}p{3cm}p{3cm}}
\hline
Method & Material Science & Engineering & Medicine \\
\hline
Markov models & \cite{Silva2019,Gassner2021205} & \cite{Huang20152061,Timilsina20234243,Streb2023,Yamaguchi2023,Park2024,Long2024,Liu202381} & \cite{Nakazawa2018343,Li2023a,Eremina2022,Ayala2018,Ohnishi2016,Hill201745,Lai2021,Tian2021} \\
Hidden Markov Models & \cite{Yang2022b,Mashshay2023,Inglese2016808,Lin202137,Goushegir2020,Gupta2023,Nepal20131060} & \cite{Peng20212105,Erdinc2009383,Volz2022,Singh20172481,Liu2012336,Preis2023,Szczerska2023,Kareem2022} & \cite{Gredic2023,Calciolari2018430,Eremina2022,Zheng2023,Kim20211031,Wang202491,Hwang2017,Tunc200424} \\
Nonparametric Bayesian Time Series Modeling & \cite{Hu2023,Szabelski2022,Kaewmala2023,AmayaGomez201926,Meng2021,Ueda2020} & \cite{Zhao2023,Chen2021,Zhao2024,Haque2023,Kulkarni2009,Huang20152061,Shchurov2021,Stroe2018517,Karunathilake2023,Abe2019,Sacramento202041,Qu2022441,Su2023,Reniers2019A3189,Nicolai2016153,Yu20061840} & \cite{Lee20162994,Elmounedi2023,Famouri2020,Li2021,Shuai2021490,Snuggs2019,Hsiao2023,Rashid2018,Song2016114} \\
\hline
\end{tabular}
\end{table}

\subsubsection{Machine Learning usage}

Table \ref{tab:methodsML} shows the use of different machine learning techniques in three disciplines: material science, power engineering, and medicine.

Table \ref{tab:methodsML} showcases articles in the field of Medicine that focus on specific techniques used in degradation models with statistical inference. Degradation models play a significant role in understanding disease progression, treatment effectiveness, and the performance of medical devices. The articles listed in Table \ref{tab:methodsML} cover a wide range of medical applications, including drug delivery systems, implantable devices, and disease prognosis. By applying statistical inference techniques to degradation data, researchers gain valuable insights into the effectiveness of treatments, the longevity of medical devices, and the progression of diseases. 

\begin{table}[htbp]
\centering
\caption{Method used for degradation modelling with machine learning}
\label{tab:methodsML}
\begin{tabular}{p{3cm}p{3cm}p{3cm}p{3cm}}
\hline
Method & Material Science & Engineering & Medicine \\
\hline
Supervised Learning & \cite{Jayakumar2021,RodriguezBravo2019394,Liu2023,Hosseini2018,Inglese2016808,Dalmau2018167,Chunsheng2017,DUrso2024} & \cite{Osara2019,Fang202332884,Preis2023,Yan2021,Oria20236539,Yilmaz2023,Kareem2022,Yit2023588,Kohtz2024} & \cite{Tanino2023,Sheng201867,Lee20162994,Kong2008327,Famouri2020,Baykal2023111,Li2021,Tunc200424} \\
Deep Learning & \cite{Wu2024,Pan2023,AlMahdi2023604,Chaudhary2023} & \cite{Shi2023,Su2024,Yit2023588,Li2024,Chen2023b,Li202239,Timilsina20234243,Jeong2023,Chen2021,Fan2021} & \cite{Calciolari2018430,Kim20211031,Khan202232750,RiveroRios2016238} \\
Reinforcement Learning & \cite{Zheng2024,Badet2020,Yang2022a,Mashshay2023,Liu2022,Elahi202330} & \cite{Mitra2023,Kathribail2023305,Schiffmacher2021895,Costa2022,Alam20135592,Narale2019,Szczerska2023,Lu2023} & \cite{Ji2021261,Nakazawa2018343,Duan2024,Li2023,Godinho2021,Vernizzi2020,Elmounedi2023,Mokeem2023,Hill201745} \\
Cluster analysis & \cite{Nguyen2024,Pan2023,Goushegir2020,Xiang2011768,Kudo201531,AfsarDizaj20225374,Kropka201514,Li2022,Gassner2021205} & \cite{Chaleshtori2024,Long2024,Volz2022,Abe2019,Thornton2024,Kim20221002,Jain2024,Bai2023,Han2022} & \cite{Duan2024,Tian2021,Betancourt2019,Khan202232750,Mokeem2023,Baykal2023111} \\
\hline
\end{tabular}
\end{table}

Supervised Learning is applicable in Material Science, Engineering, and Medicine. It involves training a model using labeled data to make predictions or classify new data points.
Deep Learning is more applicable in Engineering compared to Material Science and Medicine. It is a subset of machine learning that focuses on training deep neural networks to learn hierarchical representations of data.
Reinforcement Learning is applicable in Engineering and Medicine, with higher relevance in Medicine. It involves training an agent to make decisions in an environment to maximize a reward signal.
Cluster Analysis is more applicable in Material Science compared to Engineering and Medicine. It is an unsupervised learning technique used to group similar data points together based on their inherent characteristics.

\subsection{Hybrid methods}

\subsubsection{Statistical Inference and Dynamic Prediction}
\begin{figure}[!tbp]
    \centering
    \subfloat[Regression Analysis\label{fig:SIDP_RA2}]{
        \includegraphics[width=.48\linewidth]{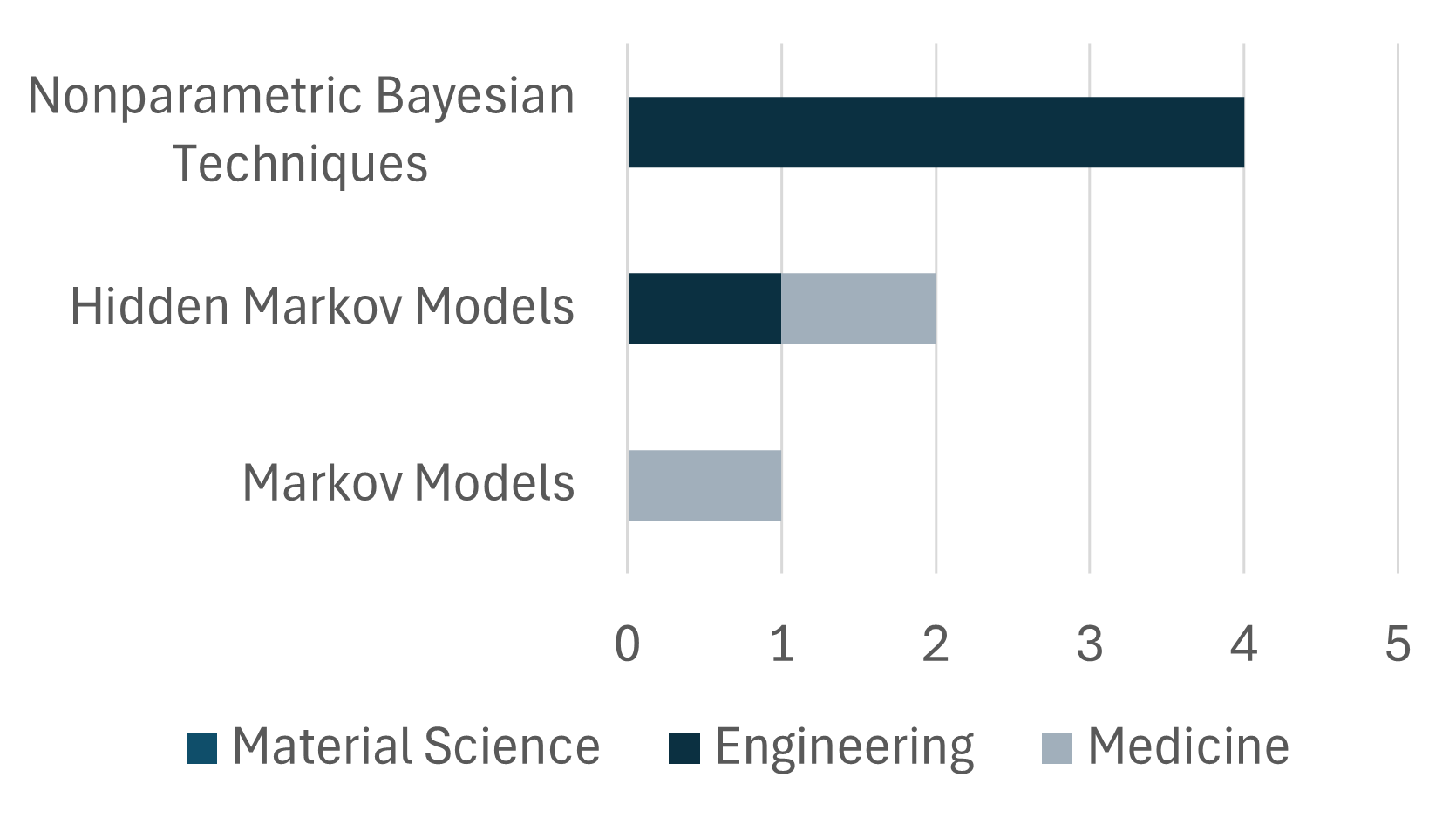}
    }\hfill
    \subfloat[Bayesian statistics\label{fig:SIDP_BT2}]{
        \includegraphics[width=.48\linewidth]{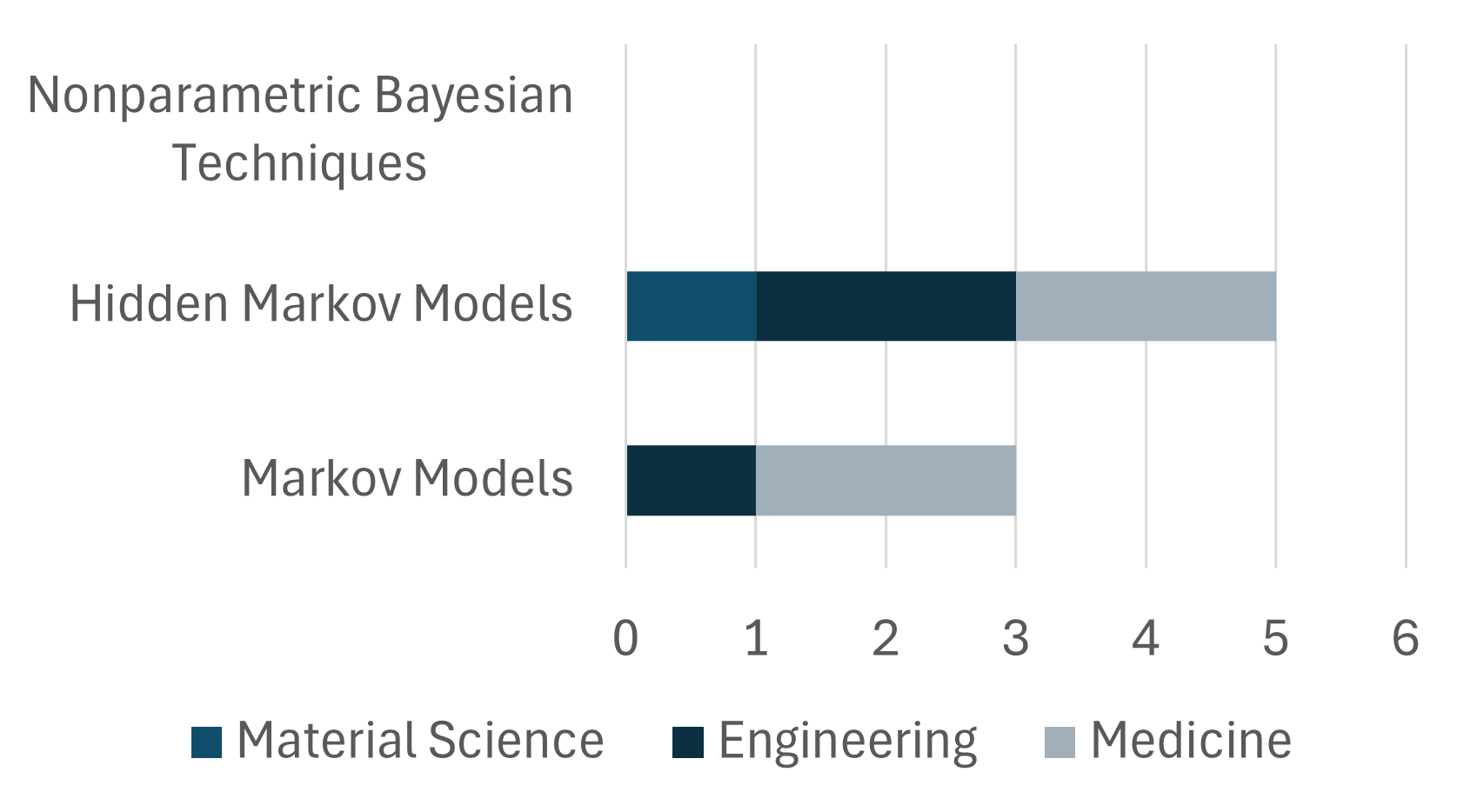}
    }\\[0.6em]
    \subfloat[Factor Analysis\label{fig:SIDP_FA2}]{
        \includegraphics[width=.48\linewidth]{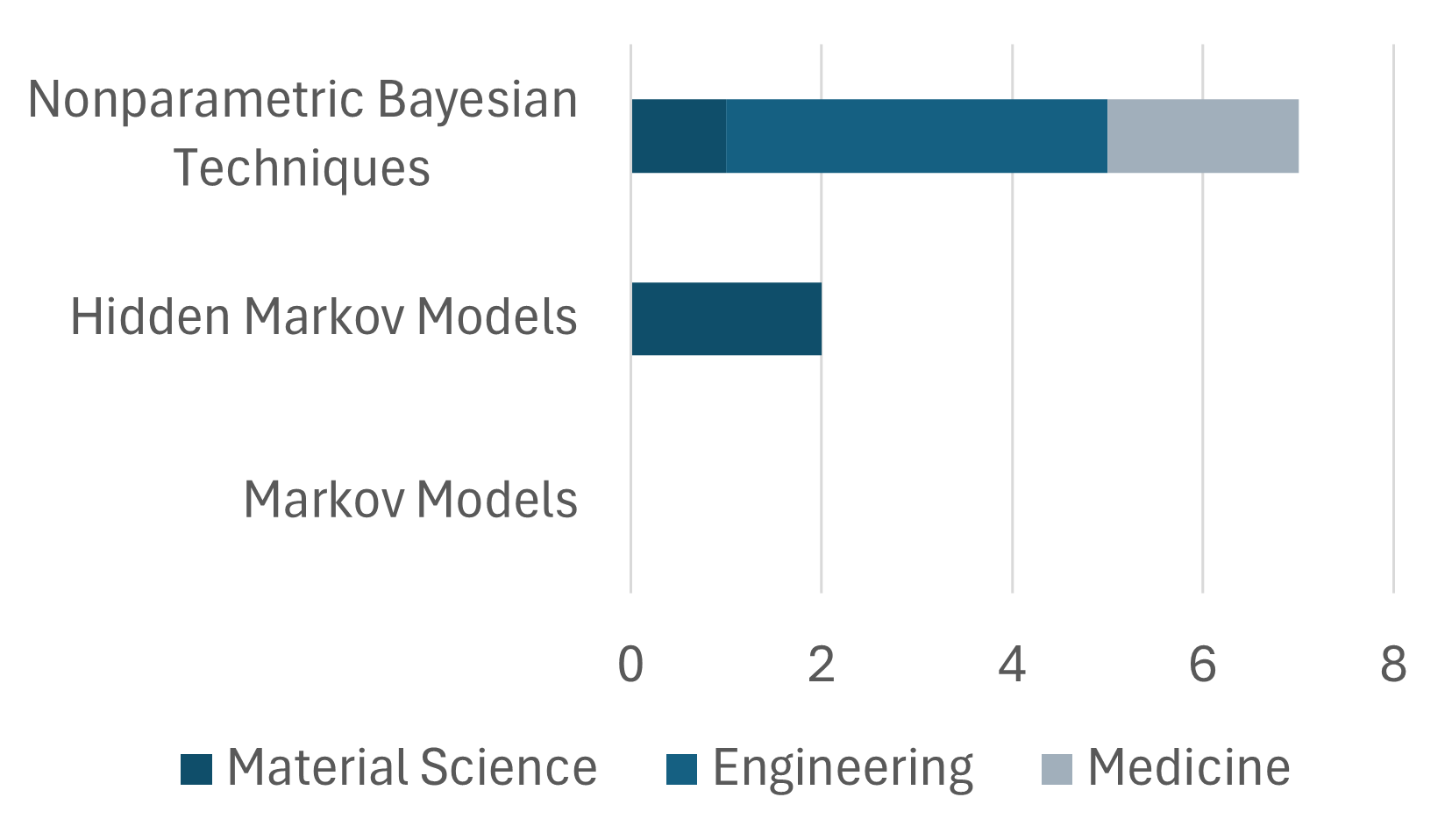}
    }\hfill
    \subfloat[MCMC and probabilistic programming\label{fig:SIDP_MCMC2}]{
        \includegraphics[width=.48\linewidth]{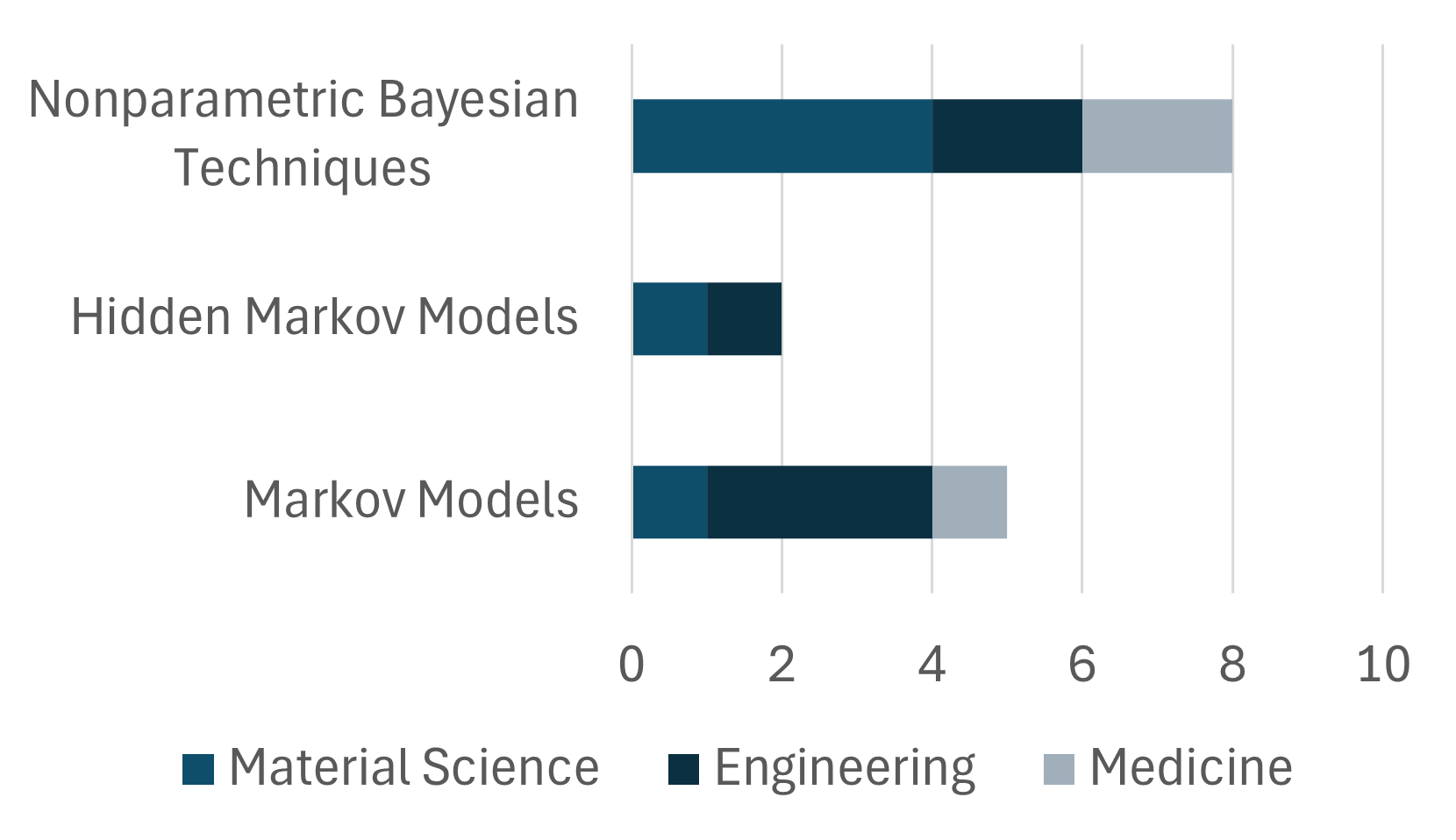}
    }
    \caption{Combination of statistical inference and dynamic prediction methods}
    \label{fig:SIDP}
\end{figure}

Figure \ref{fig:SIDP} presents combinations of methods from statistical inference and dynamic prediction. It demonstrates the usage of MCMC in material science, power engineering, and medicine for parameter estimation, exploring distributions, and Bayesian models. MCMC simulations are valuable for simulating complex systems and addressing uncertainty in diverse domains. Nonparametric Bayesian modeling is important in material science, while in power engineering and medicine, MCMC is used for uncertainty quantification and computational modeling.


There is no specific relevance mentioned for the combination of Regression Analysis and Markov Models in any of the studied disciplines.
The combination of Bayesian statistics and Hidden Markov Models is applicable in all three disciplines, with higher relevance in Engineering and Medicine. This suggests that these techniques can be effectively used together to analyze and model complex systems.
The combination of Marcov Chain Monte Carlo and Nonparametric Bayesian Techniques is more applicable in Material Science and Medicine compared to Engineering. This indicates the potential for utilizing these techniques together in solving problems related to sampling from complex probability distributions and analyzing time series data. 

\subsubsection{Statistical Inference and Machine Learning}

\begin{figure}[!tbp]
    \centering

\subfloat[Regression Analysis]{
\includegraphics[width=0.48\linewidth]
{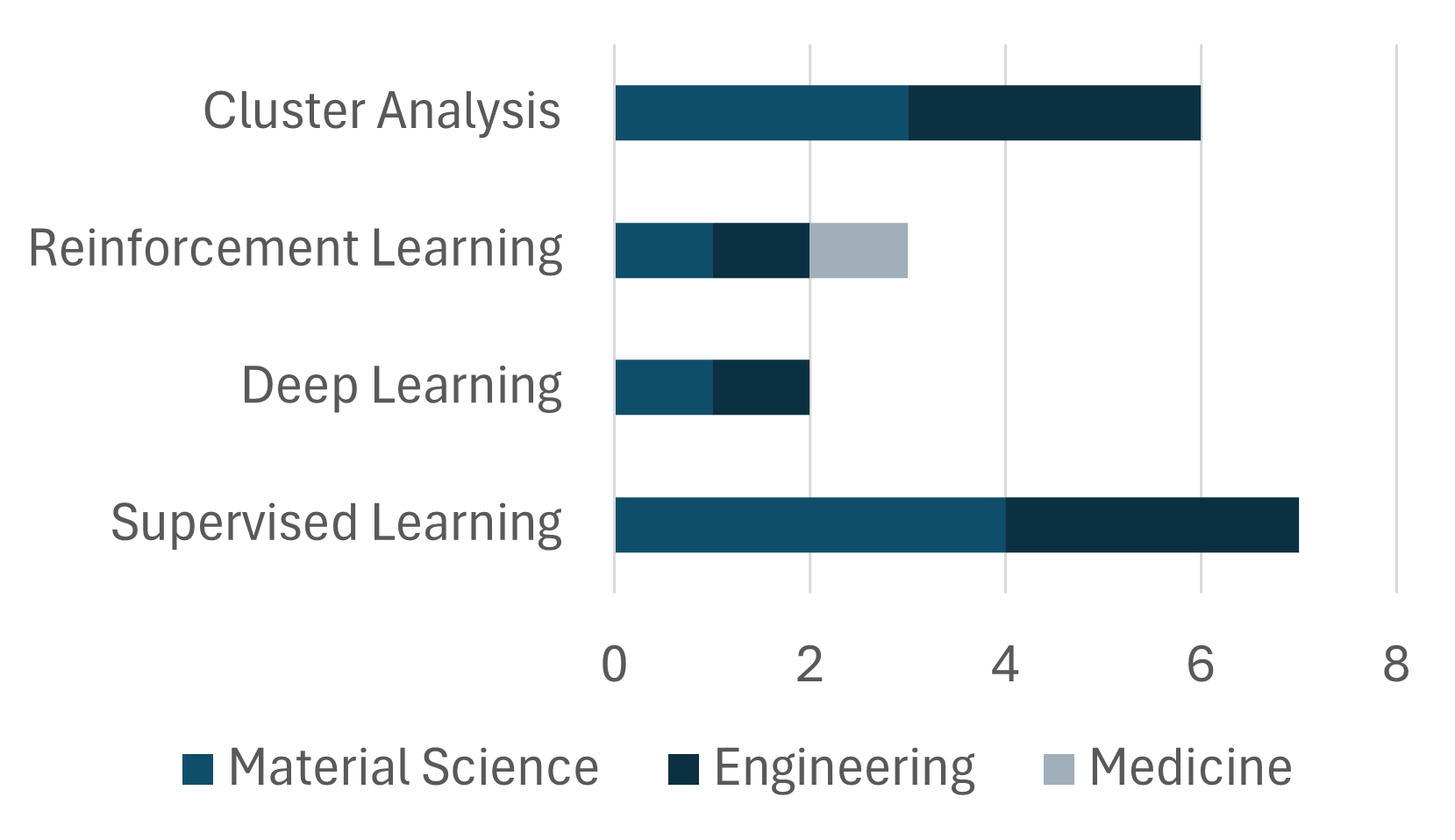}
    \label{fig:SIML_RA}
}    ~ 
\subfloat[Bayesian statistics]{
\includegraphics[width=0.48\linewidth]{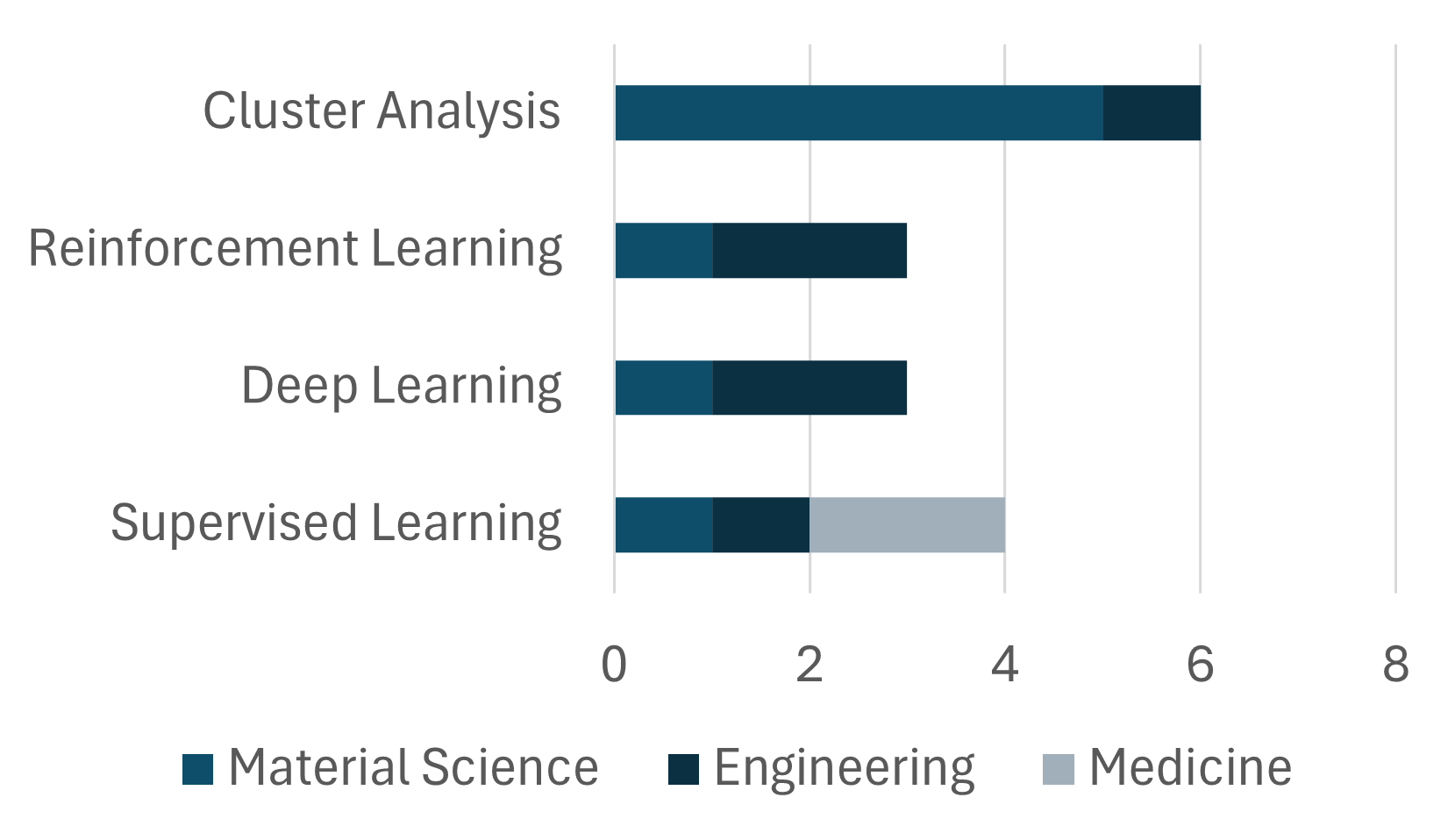}
    \label{fig:SIML_BT}
}
\\
\subfloat[Factor Analysis]{
\includegraphics[width=0.48\linewidth]{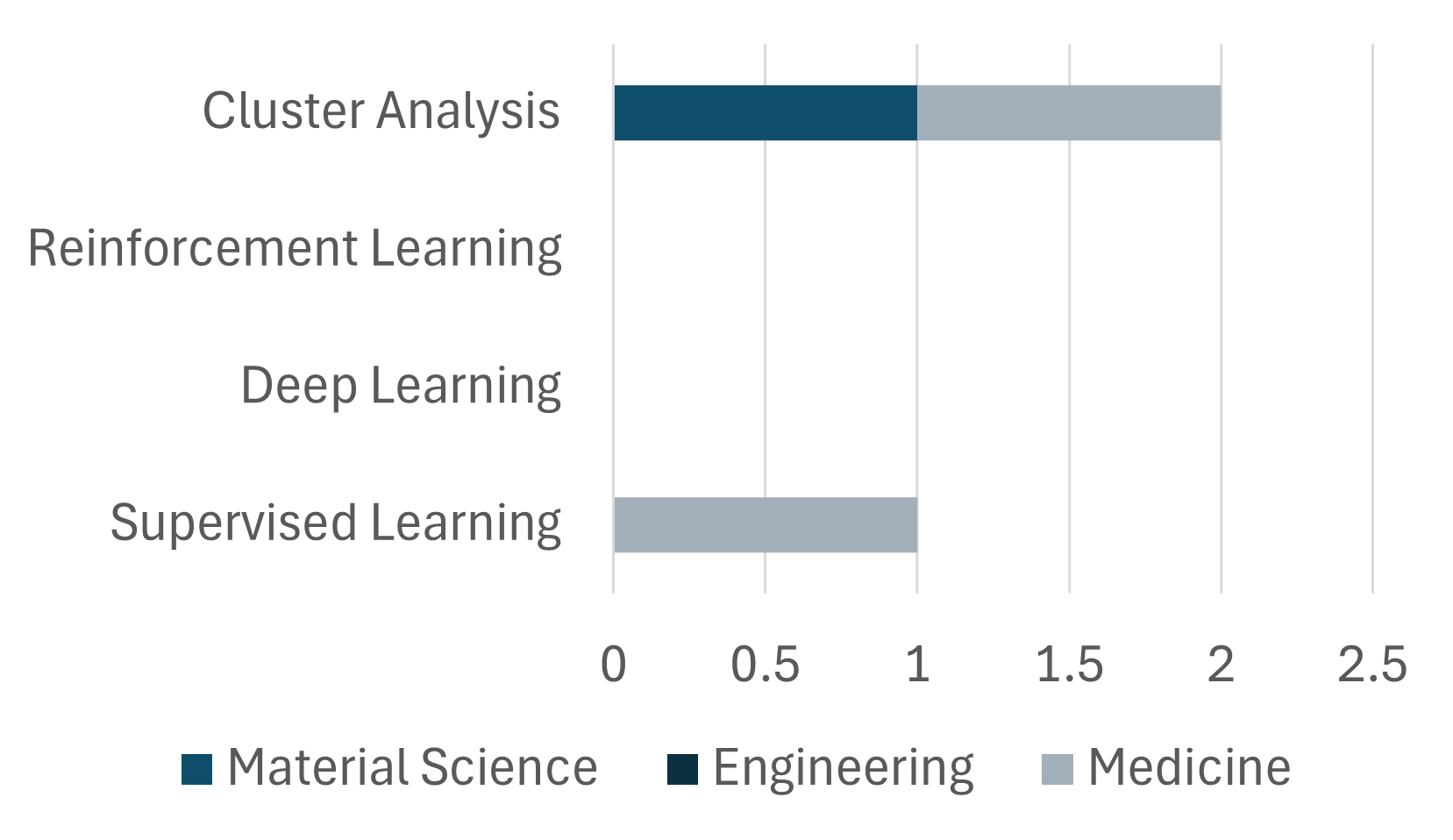}
    \label{fig:SIML_FA}
}
  ~
\subfloat[MCMC and probabilistic programming]{
\includegraphics[width=0.48\linewidth]{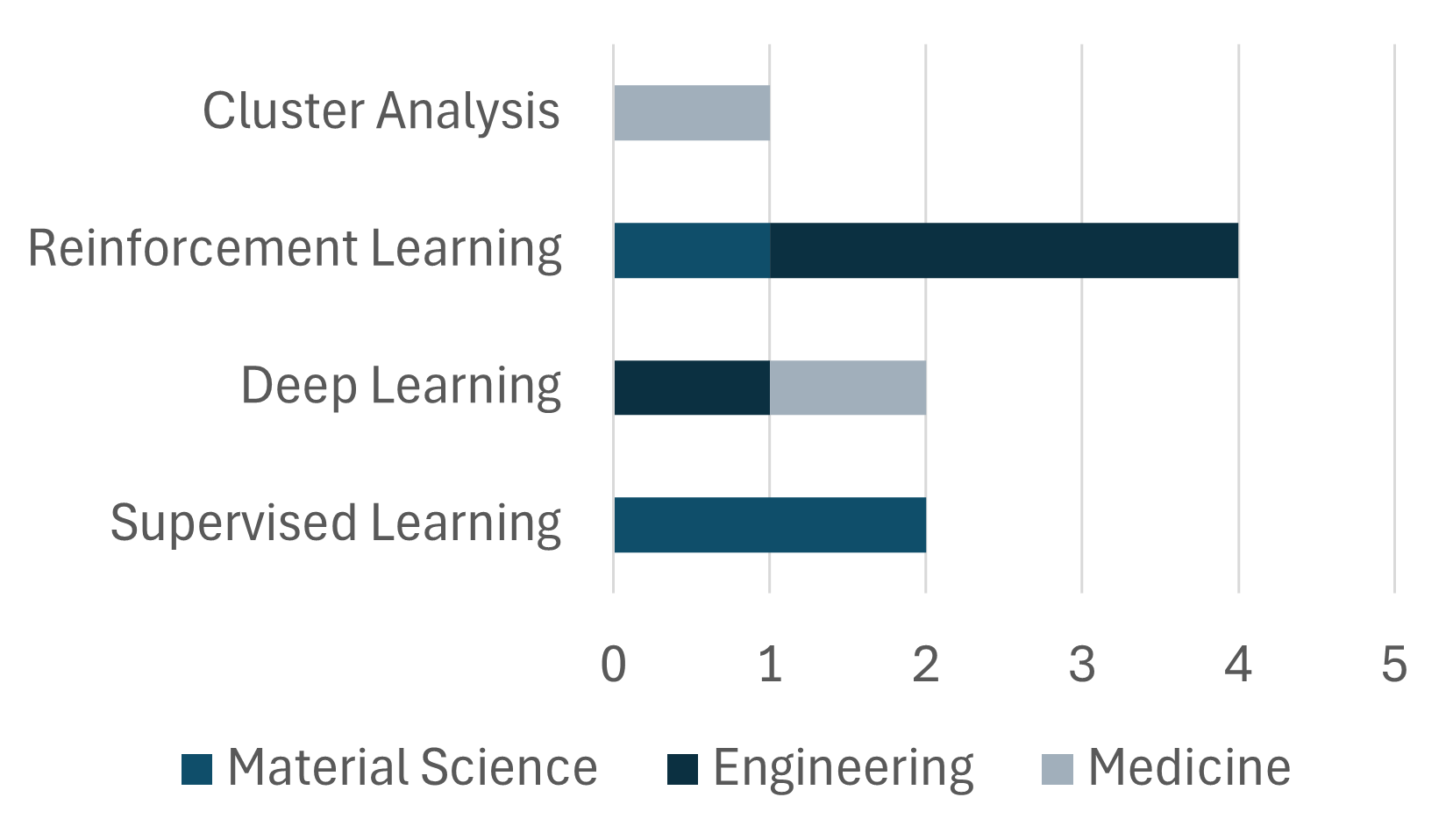}
    \label{fig:SIML_MCMC}}
    \caption{Combination of statistical inference and machine learning}
\label{fig:SIML}
\end{figure}

Figure \ref{fig:SIML} presents combinations of techniques from the area of statistical inference methods and machine learning techniques. 

The combination of Markov Chain Monte Carlo and Probabilistic Programming demonstrates varying relevance across the disciplines. It is more applicable in Engineering for Deep Learning and Reinforcement Learning techniques, while in Material Science, it shows particular relevance for Supervised Learning. Medicine exhibits limited relevance mentioned for Cluster Analysis in this combination.
The combination of Bayesian statistics and Supervised Learning is applicable in all three disciplines, with higher relevance in Material Science and Medicine. It finds applications in analyzing and modeling data in the context of supervised learning tasks.
The combination of Regression Analysis and Deep Learning is applicable in Material Science and Engineering, with no specific relevance mentioned in Medicine. This suggests the potential integration of regression analysis with deep learning techniques for data analysis.

\subsubsection{Dynamic Prediction and Machine Learning}

\begin{figure}[!tbp]
    \centering
    \subfloat[Supervised Learning\label{fig:DPML_SL}]{
        \includegraphics[width=.48\linewidth]{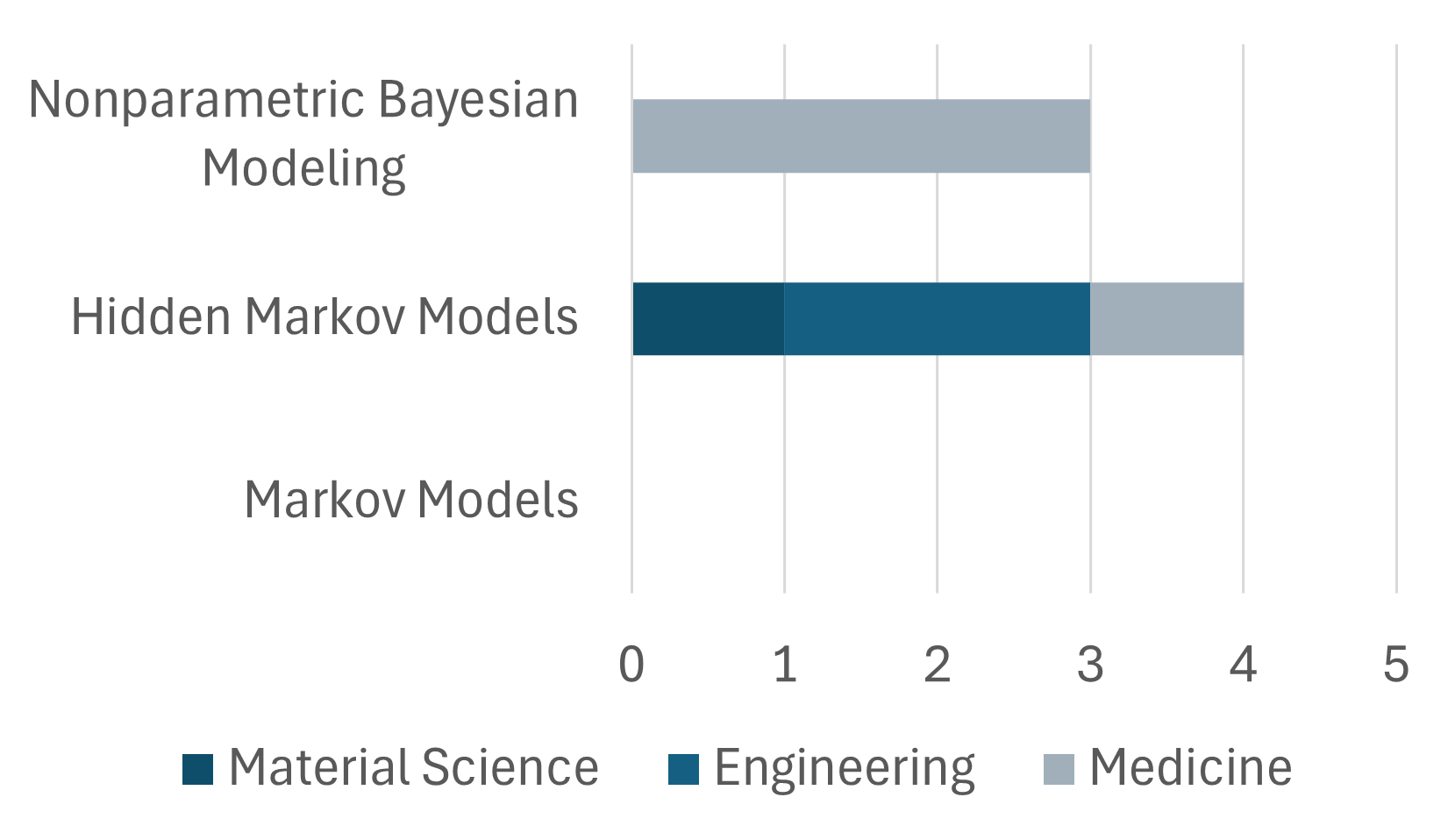}
    }\hfill
    \subfloat[Deep Learning\label{fig:DPML_DL}]{
        \includegraphics[width=.48\linewidth]{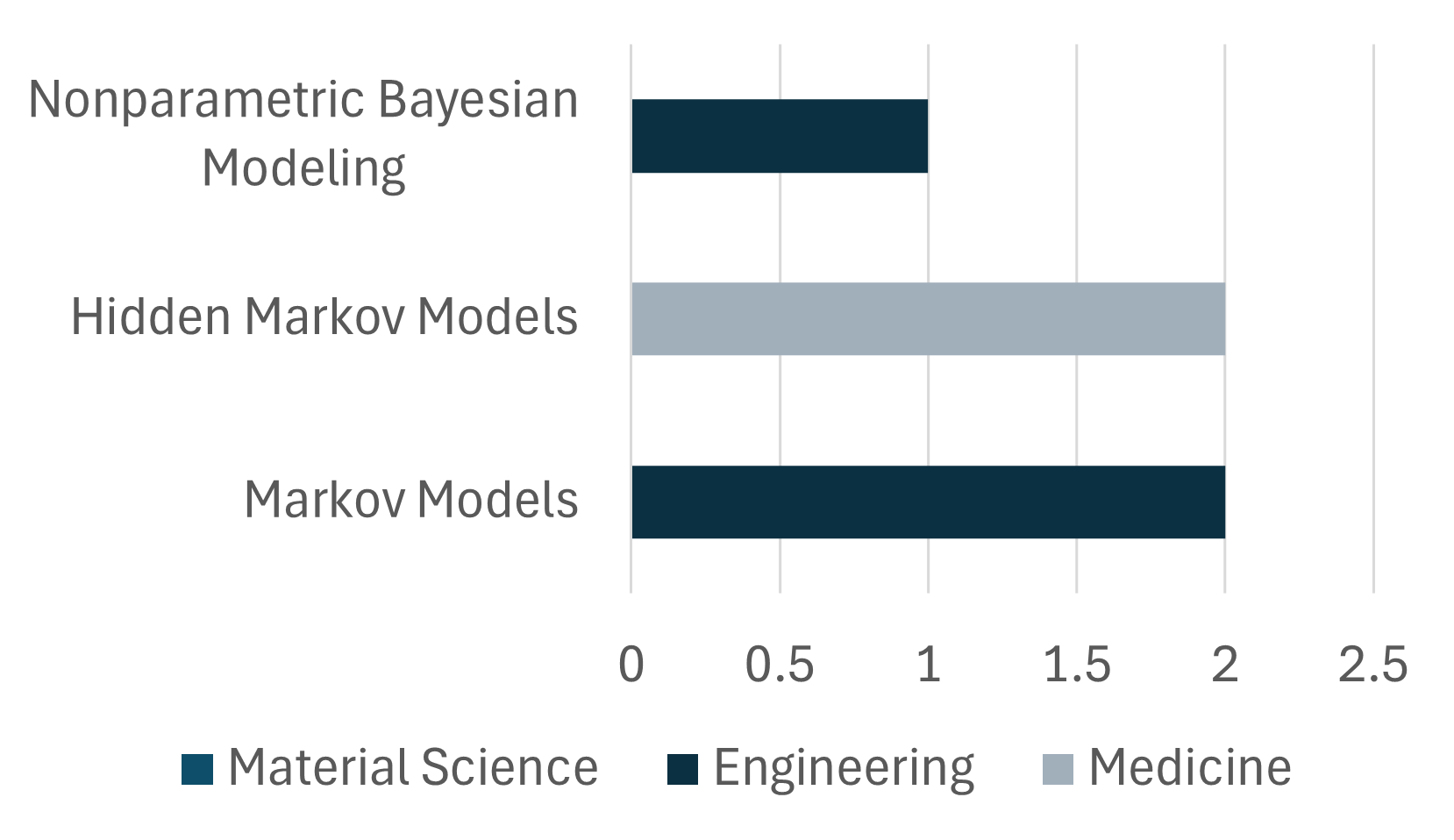}
    }\\[0.6em]
    \subfloat[Reinforcement Learning\label{fig:DPML_RL}]{
        \includegraphics[width=.48\linewidth]{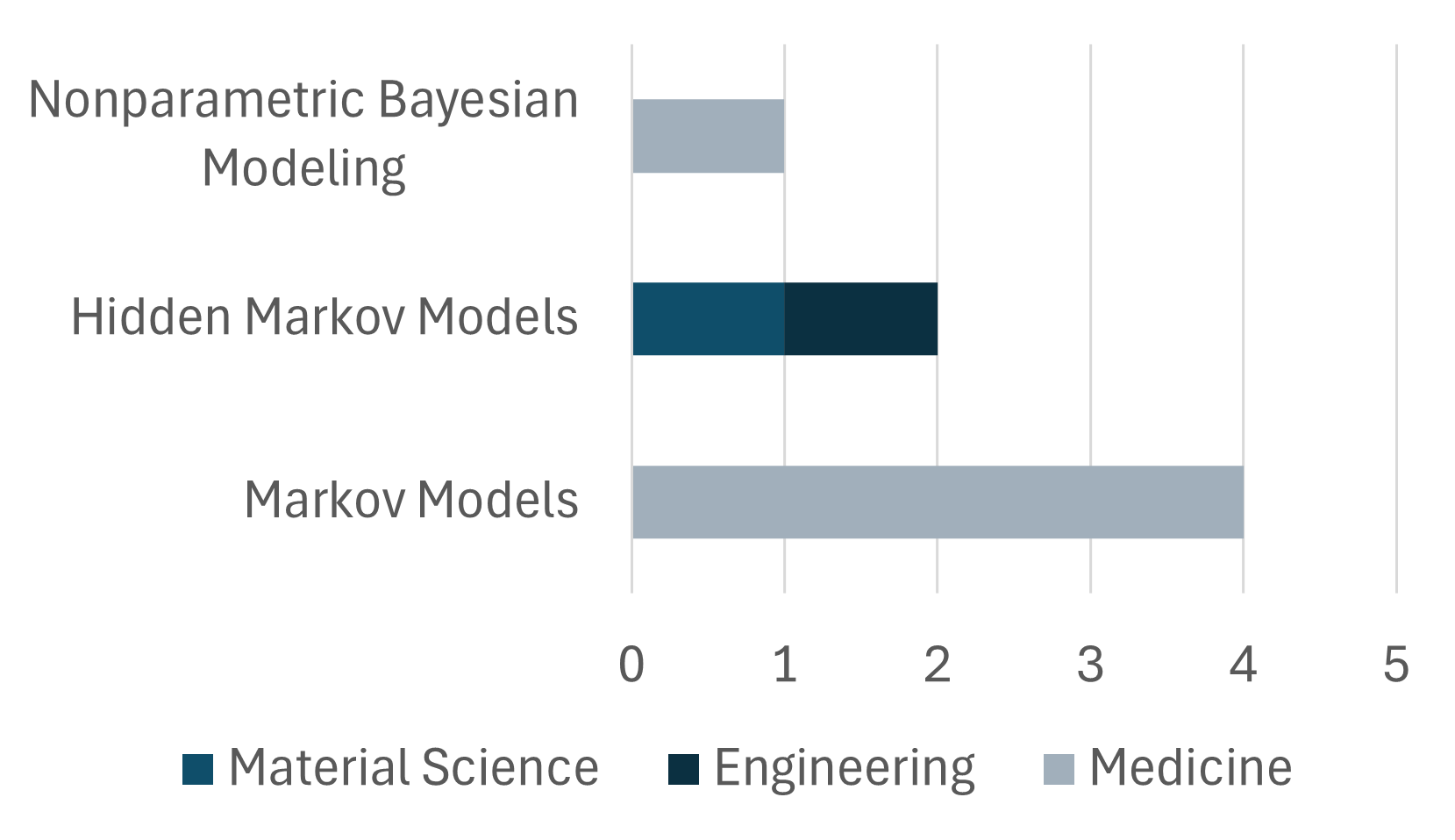}
    }\hfill
    \subfloat[Cluster Analysis\label{fig:DPML_CA}]{
        \includegraphics[width=.48\linewidth]{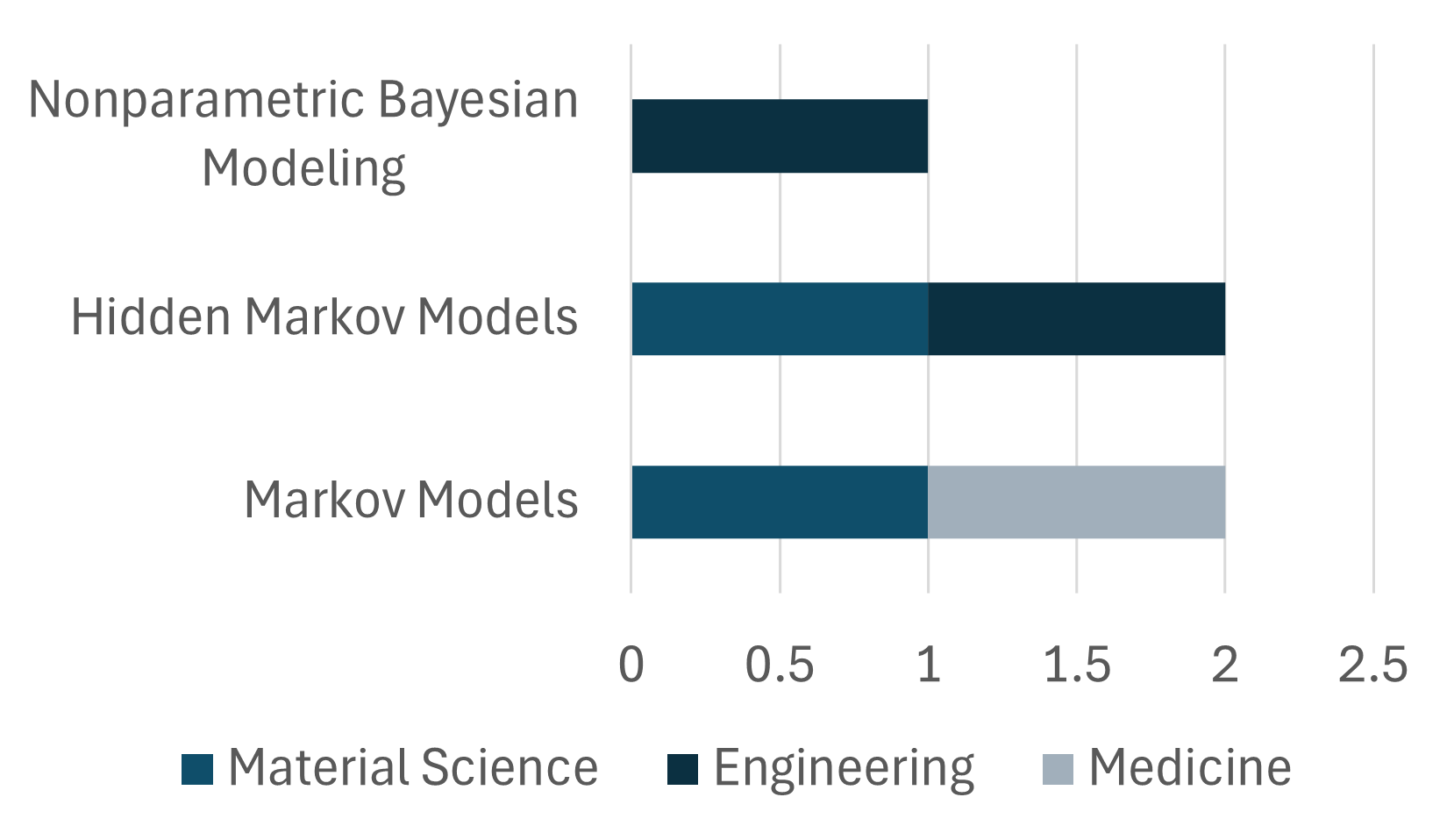}
    }
    \caption{Combination of dynamic prediction and machine learning}
    \label{fig:DPML}
\end{figure}

Figure \ref{fig:DPML} presents the combination of machine learning techniques and dynamic prediction in scientific fields. Figure \ref{fig:DPML} showcases the relevance of different combinations of methods in various disciplines, including Material Science, Engineering, and Medicine.

The combination of Supervised Learning and Hidden Markov Models is applicable in all three disciplines, with higher relevance in Medicine. This indicates the potential for utilizing these techniques together to analyze and model data in the context of supervised learning tasks.
The combination of Deep Learning and Hidden Markov Models is applicable in Engineering and Medicine, with no specific relevance mentioned in Material Science. This suggests the potential integration of deep learning techniques with hidden Markov models for data analysis in these disciplines.
The combination of Reinforcement Learning and Markov Models is particularly applicable in Medicine, with higher relevance compared to Material Science and Engineering. This highlights the potential for utilizing reinforcement learning techniques in analyzing and optimizing complex systems within the medical field.
The combination of Cluster Analysis and Markov Models is applicable in Material Science, indicating the potential for utilizing cluster analysis techniques in analyzing and identifying patterns in data within this discipline.

\section{Discussion and challenges}

\subsection{Discussion}
Table \ref{tbl:PlusMinus} presents an overview of the advantages and disadvantages of the methods for degradation modelling in this paper.

\begin{table}[!tbp]
\begin{adjustwidth}{-0.1\extralength}{0cm}

\caption{Comparison of strengths and weaknesses of each class of methods. Statistical approaches are more interpretable, but can be more costly to compute, however machine learning when provided enough data has great predictive performance. }
\centering
\begin{tabular}{@{}llll@{}}
\toprule
                    & Methods                                                                       & Advantages                                                                                                 & Disadvantages                                                                                                           \\ \midrule
\multirow{4}{*}{\STAB{\rotatebox[origin=c]{90}{Statistical inference}}} & Regression analysis                                                           & \begin{tabular}[c]{@{}l@{}}Fast computation,\\ Easily interpretable\end{tabular}                           & \begin{tabular}[c]{@{}l@{}}Linear relationships,\\ Strong assumptions\end{tabular}                                      \\
                    & Factor analysis                                                               & Interpretability                                                                                           & \begin{tabular}[c]{@{}l@{}}Highest expert \\ knowledge requirements\end{tabular}                                        \\
                    & \begin{tabular}[c]{@{}l@{}}MCMC \&\\ probabilistic \\ programming\end{tabular} & \begin{tabular}[c]{@{}l@{}}Specification of\\ complicated models,\end{tabular}                             & \begin{tabular}[c]{@{}l@{}}Mostly Bayesian\\ applications,\\ Computational costs\end{tabular}                           \\
                    & Bayesian statistics                                                           & \begin{tabular}[c]{@{}l@{}}Most flexibility,\\ interpretability\end{tabular}                               & \begin{tabular}[c]{@{}l@{}}Computational problems, \\ untraceablity of direct\\ analytical formulas\end{tabular}        \\ \midrule
\multirow{3}{*}{\STAB{\rotatebox[origin=c]{90}{Dynamic prediction}}} & Markov models                                                                 & \begin{tabular}[c]{@{}l@{}}Widely understood,\\ Good short \\ term predictions\end{tabular}                & Limited long term accuracy                                                                                              \\
                    & Hidden Markov models                                                          & \begin{tabular}[c]{@{}l@{}}Improved prediction \\ quality, more flexibility\end{tabular}                   & Difficult interpretability                                                                                              \\
                    & Bayesian time series                                                          & \begin{tabular}[c]{@{}l@{}}Ease of application \\ (esp. Prophet), \\decomposition of effects\end{tabular}                             & Computational costs                                                                                                     \\
                    &&&\\
                    \midrule
\multirow{4}{*}{\rotatebox[origin=c]{90}{Machine learning\hspace{1cm}}} & Supervised learning                                                           & Flexibility, wide adoption                                                                                 & \begin{tabular}[c]{@{}l@{}}Difficult interpretability, \\ large dataset requirements\end{tabular}                       \\
                    & Deep learning                                                                 & \begin{tabular}[c]{@{}l@{}}Flexibility, predictive \\ performance\end{tabular}                             & \begin{tabular}[c]{@{}l@{}} High dataset \\ requirements, no interpretability,\\ computational costs\end{tabular} \\
                    & Reinforcement learning                                                        & Very promising results                                                                                     & \begin{tabular}[c]{@{}l@{}} High dataset \\ requirements, no interpretability,\\ computational costs\end{tabular} \\
                    & Cluster analysis                                                              & \begin{tabular}[c]{@{}l@{}}Limited expert knowledge\\ required, detection of\\ atypical cases\end{tabular} & \begin{tabular}[c]{@{}l@{}}Data requirements,\\ Problems with traceability\end{tabular}                                 \\ \bottomrule
\end{tabular}
\label{tbl:PlusMinus}
\end{adjustwidth}
\end{table}

\subsection{Open challenges}
Researchers may explore new modeling techniques which include investigating physics-based models, data-driven models, and hybrid models that combine different approaches to improve the accuracy and reliability of degradation predictions. Additionally, incorporating uncertainty analysis techniques, such as hierarchical nature of degradation processes.

Managing missing data is another critical aspect that needs attention. Strategies such as imputation techniques or robust estimation methods can be employed to handle missing data and ensure accurate model predictions. Furthermore, exploring data fusion and integration techniques, which integrate multiple sources of data such as sensor data, historical records, and expert knowledge, can enhance the understanding and prediction of degradation processes.

Researchers may establish benchmark datasets that are standardized and representative of different degradation scenarios to assess and compare the performance of degradation models. It is essential to define appropriate evaluation metrics that capture the unique characteristics of degradation modeling, considering factors such as prediction accuracy, robustness to uncertainties, and computational efficiency.

Researchers may develop models that capture the dynamic nature of degradation processes to make degradation modeling more applicable in real-world scenarios. These models may consider factors such as aging, maintenance interventions, and changes in operating conditions. Additionally, addressing scalability issues is crucial, as degradation modeling often deals with large-scale systems. 

Exploring scalable techniques that can handle numerous components or spatially distributed degradation processes is necessary. Real-time degradation modeling techniques may also be investigated to provide timely predictions and support proactive decision-making in various industries. Through this action researchers can contribute to the continuous improvement and application of degradation modeling, leading to enhanced understanding, prediction, and management of degradation processes in diverse domains.

 \section{Conclusions}
In conclusion, the literature review on degradation modeling methods, including regression analysis, factor analysis, cluster analysis, Markov Chain Monte Carlo with probabilistic programming, Bayesian statistics, hidden Markov models, nonparametric Bayesian modeling of time series, supervised learning, and deep learning, provides valuable insights into the diverse applications and approaches in the field. The review encompasses a comprehensive exploration of degradation modeling across various domains, including material science, power engineering, and medicine.

The review highlights the significance of understanding degradation patterns, optimizing maintenance strategies, and improving overall process efficiency in different domains. It emphasizes the critical role of degradation modeling in analyzing degradation mechanisms, optimizing treatment strategies, and predicting disease progression to improve patient outcomes in the field of Medicine. Additionally, the review underscores the importance of employing statistical inference, dynamic prediction methods, and machine learning techniques in degradation modeling across different disciplines.

Furthermore, the review sheds light on the potential of hybrid modeling techniques, such as the combination of statistical inference and dynamic prediction, statistical inference and machine learning, and dynamic prediction and machine learning, to enhance the understanding and prediction of degradation processes.

The review provides a foundation for understanding the common themes and methods employed in degradation modeling, as evidenced by the overview of various degradation models, their references, methods, and applications. It also offers valuable insights into the preferences and popularity of specific techniques in particular fields, as demonstrated by the number of applications of individual techniques from the Dynamic prediction methods method in each discipline.

Overall, the review serves as a comprehensive resource for researchers and practitioners in the field of degradation modeling, offering a rich understanding of the methods, applications, and implications across diverse domains.

\vspace{6pt} 





\authorcontributions{Conceptualization, AJK and JB; Methodology (search strategy, eligibility criteria, quality appraisal), AJK and JB; Investigation (database searching, screening, full‐text review), AJK; Data Curation (reference management, data extraction), AJK; Formal Analysis (qualitative synthesis), AJK and JB; Validation (duplicate screening/extraction checks), JB; Visualization (PRISMA diagram, summary tables), AJK; Writing—original draft preparation, AJK; Writing—review and editing, AJK and JB; Supervision, JB; Project administration, JB; Funding acquisition, JB. All authors have read and agreed to the published version of the manuscript.
}

\funding{Work of Anna Jarosz and Jerzy Baranowski partially realised in the scope of project titled ''Process Fault Prediction and Detection''. Project was financed by The National Science Centre on the base of decision no. UMO-2021/41/B/ST7/03851. Part of work was funded by AGH’s Research University Excellence Initiative under project “DUDU - Diagnostyka Uszkodzeń i Degradacji Urzadzeń''. }

\institutionalreview{Not applicable}

\informedconsent{Not applicable}

\dataavailability{Data is contained within the article or supplementary material}

\acknowledgments{Authors would like to thank Dr Marta Zagorowska for her great help in manuscript consultation. Authors would also like to express their gratitude to Prof. Xiaosong Du for extending invitation to submit. 

During the preparation of this manuscript, authors used ChatGPT 5 for the purposes of style verification and refinement. The authors have reviewed and edited the output and take full responsibility for the content of this publication.}

\conflictsofinterest{The authors declare no conflicts of interest.} 



\abbreviations{Abbreviations}{
The following abbreviations are used in this manuscript:
\\

\noindent 
\begin{tabular}{@{}ll}
ISO & International Organization for Standardization\\
OLS & Ordinary Least Squares\\
AIC & Akaike Information Criterion\\
MCMC & Markov Chain Monte Carlo\\
HMC & Hamiltonian Monte Carlo\\
HMM & Hidden Markov Model\\
PPL & Probabilistic Programming Language\\
GP & Gaussian Process\\
DP & Dirichlet Process\\
MAP & Maximum a Posteriori (estimation)\\
EM & Expectation–Maximization (algorithm)\\
MDP & Markov Decision Process\\
CNN & Convolutional Neural Network\\
RNN & Recurrent Neural Network\\
LSTM & Long Short-Term Memory (network)\\
RUL & Remaining Useful Life\\
VC & Vapnik–Chervonenkis (dimension)\\
H$_2$S & Hydrogen sulfide
\end{tabular}
}

\isPreprints{}{
\begin{adjustwidth}{-\extralength}{0cm}
} 

\reftitle{References}

\PublishersNote{}
\isPreprints{}{
\end{adjustwidth}
} 
\end{document}